\documentclass[10pt]{article}
\usepackage{epsf}

\usepackage{epsfig,graphics}
\usepackage {graphicx}
\usepackage {epsfig}
\usepackage {subfigure}
\usepackage {tabularx} 
\usepackage{rotate}	
\usepackage{slashed}

\usepackage{amsmath}
\usepackage{amsfonts}
\usepackage{amssymb}
\usepackage{graphicx}
\usepackage{cite}
\usepackage{float}

\usepackage{fancyhdr}
\usepackage{hyperref}

 \usepackage{tikz-feynman} 
\usepackage{lettrine}
\usepackage{graphicx}
\usepackage{relsize}

\newcommand{\bmat}{\left(\begin{array}}
	\newcommand{\emat}{\end{array}\right)}

\def\yzero{\smash{\hbox{$y\kern-4pt\raise1pt\hbox{${}^\circ$}$}}}

\def\beq{\begin{equation}}
\def\eeq{\end{equation}}
\def\beqa{\begin{eqnarray}}
\def\eeqa{\end{eqnarray}}

\def\-{\hphantom{-}}

\def\s2{\frac{1}{\sqrt2}}

\def\beq{\begin{equation}}
\def\eeq{\end{equation}}
\def\beqa{\begin{eqnarray}}
\def\eeqa{\end{eqnarray}}

\def\IF{\relax{\rm I\kern-.18em F}}
\def\II{\relax{\rm I\kern-.18em I}}
\def\IP{\relax{\rm I\kern-.18em P}}
\def\IC{\relax\hbox{\kern.25em$\inbar\kern-.3em{\rm C}$}}
\def\IR{\relax{\rm I\kern-.18em R}}

\def\Dsl{\,\raise.15ex\hbox{/}\mkern-13.5mu D} 
\def\IZ{Z\kern-.4em  Z}

\catcode`\@=11   
\newdimen\@rotdimen
\newbox\@rotbox  

\def\@vspec#1{\special{ps:#1}}
\def\@rotstart#1{\@vspec{gsave currentpoint currentpoint translate
		#1 neg exch neg exch translate}}
\def\@rotfinish{\@vspec{currentpoint grestore moveto}}
\def\@rotr#1{\@rotdimen=\ht#1\advance\@rotdimen by\dp#1%
	\hbox to\@rotdimen{\hskip\ht#1\vbox to\wd#1{\@rotstart{90 rotate}%
			\box#1\vss}\hss}\@rotfinish}
\def\@rotl#1{\@rotdimen=\ht#1\advance\@rotdimen by\dp#1%
	\hbox to\@rotdimen{\vbox to\wd#1{\vskip\wd#1\@rotstart{270 rotate}%
			\box#1\vss}\hss}\@rotfinish}%
\def\@rotu#1{\@rotdimen=\ht#1\advance\@rotdimen by\dp#1%
	\hbox to\wd#1{\hskip\wd#1\vbox to\@rotdimen{\vskip\@rotdimen
			\@rotstart{-1 dup scale}\box#1\vss}\hss}\@rotfinish}%
\def\@rotf#1{\hbox to\wd#1{\hskip\wd#1\@rotstart{-1 1 scale}%
		\box#1\hss}\@rotfinish}%
\def\rotate{\@ifnextchar[{\@rotate}{\@rotate[l]}}
\def\@rotate[#1]#2{\setbox\@rotbox=\hbox{#2}\@nameuse{@rot#1}\@rotbox}

\catcode`\@=12

\topmargin
-1.5cm
\textwidth
15.5cm
\textheight
23.5cm
\oddsidemargin
0.7cm
\evensidemargin
0.7cm

\begin{document}

	\makeatletter
	\@addtoreset{equation}{section}
	\makeatother
	\renewcommand{\theequation}{\thesection.\arabic{equation}}
	\pagestyle{empty}
	\vspace{-0.2cm}
	\rightline{ IFT-UAM/CSIC-20-59}
	\vspace{1.2cm}
	\begin{center}
		
		
		\LARGE{ Pair Production and Gravity as the Weakest Force 
			\\ [13mm]}
		
		\large{Eduardo Gonzalo  and Luis E. Ib\'a\~nez \\[6mm]}
		\small{
			Departamento de F\'{\i}sica Te\'orica
			and Instituto de F\'{\i}sica Te\'orica UAM/CSIC,\\[-0.3em]
			Universidad Aut\'onoma de Madrid,
			Cantoblanco, 28049 Madrid, Spain 
			\\[8mm]}
		\small{\bf Abstract} \\[6mm]
	\end{center}
	\begin{center}
		\begin{minipage}[h]{15.22cm}
		The Weak Gravity Conjecture (WGC)  is usually  formulated in terms of the stability of extremal black-holes or in terms of
		long distance Coulomb/Newton potentials.
		However one can think of other physical
		processes to compare the relative strength of gravity versus other forces.
		We argue for an alternative formulation in terms of particle pair production at threshold or,
		equivalently, pair annihilation at rest.
		Imposing that the production rate by
		any force mediator (photon or scalar) of pairs of charged particles be larger or equal to graviton production, we recover known
		conditions  for the $U(1)$ WGC and its extensions. Unlike other formulations though, threshold pair production is sensitive
		to short range couplings present in scalar interactions and gives rise to a Scalar WGC.
		Application to moduli scalars gives rise to 
		specific conditions on the trilinear and quartic couplings which involve first and second derivatives of the WGC particle mass
		with respect to the moduli. Some solutions saturating equations correspond to massive states behaving like
		BPS, KK and winding states which feature duality invariance and are  in agreement with the Swampland distance conjecture.
		Conditions for $N=2$ BPS states saturate our bounds and we discuss specific examples of BPS states 
		which become massless at large Kahler moduli in Type IIA  N=2, D=4 CY and orbifold compactifications.
		We study possible implications for potentials  depending on moduli only through WGC massive states.
				For some simple classes of potentials one recovers constraints somewhat similar 
		but not equivalent to a Swampland dS conjecture.

		\end{minipage}
	\end{center}
	\newpage
	\setcounter{page}{1}
	\pagestyle{plain}
	\renewcommand{\thefootnote}{\arabic{footnote}}
	\setcounter{footnote}{0}
	

	
	\tableofcontents
	
	\section{Introduction}

The first formulations of the Weak Gravity Conjecture (WGC) rested heavily on black-hole physics.
The simplest version of the $U(1)$  Weak Gravity Conjecture  \cite{swampland,WGC,distance} (see \cite{review} for a recent review and
references) may be formulated from the kinematic condition  that extremal black-holes can decay, which requires that a particle with 
charge $e$ and mass $m$ must exist such that $\sqrt{2}e\geq m/M_p$.  This may also be understood as a condition between the
strengths of  the gravitational  and the gauge interactions. The condition corresponds to imposing 
that, between two particles  with identical masses and charges, the gauge repulsion dominates, and no
bound states form. So it is reasonable to name this as the Weak Gravity Conjecture.
This has been generalized to the case of multiple $U(1)$ interactions,
which requires some refinements \cite{remmen,Heidenreich:2015nta,Heidenreich:2016aqi}. Thus e.g.
 for an extremal black-hole to decay,  it is not enough that particles with mass $m_i$ and charge $e_i$ exist with
$\sqrt{2}e_i\geq m_i/M_p$ for each $U(1)$, but instead that a certain condition involving the {\it convex hull} is met \cite{remmen}. 
Furthermore, if we insist that the constraints remain valid under dimensional reduction, string theory examples have shown us that 
there must exist  a sublattice (or a tower)  of infinite superextremal massive charged particles verifying the appropriate generalized version 
of the constraints \cite{Heidenreich:2015nta,Heidenreich:2016aqi,montero,Andriolo:2018lvp}.
 These generalized versions of the WGC for multiple $U(1)$'s have  passed by now a number of tests within the context of string theory.

The situation becomes more complicated in the presence of scalar couplings. Scalar couplings do not carry in general a conserved
charge and the most naive arguments based on extremal black-hole stability do not directly apply. Furthermore, the
question arises whether the Swampland conditions have to do only with black-hole physics or rather with a fundamental general 
principle that gravity is always the weakest force. This would imply the wanted instability 
of extremal black-holes but it may also lead to further constraints on different systems other than black-holes.
As we said, for $U(1)$ interactions and in the absence of scalar fields, imposing that long range Coulomb forces dominate over Newton attraction gives
equivalent results than instability of extremal black-holes \cite{WGC,repulsive}.
However, if gravity is the
weakest force, the condition should apply not only to gauge couplings but also to scalar and Yukawa couplings. In particular, $d=4$ 
quartic scalar interactions are short-range and such kind of arguments based on long range forces would yield
no information about them. Moreover, since a higher dimensional graviton gives rise to lower dimensional scalar fields,
 if the principles behind the WGC are to apply in any dimension, then some form of a scalar WGC (SWGC) is expected to exist.

In order to compare the strength of gravity with other interactions  we should evaluate amplitudes or rates for some
kinematic configuration and fixed specific
momenta. In the case at hand there are essentially two ways to evaluate these rates at tree level 1) Through diagrams
involving one propagator of the considered massless mediators (photon, graviton, moduli) and 2) Through diagrams
involving the exchange of a charged massive test particle (e.g. a BPS state).  The first possibility involves only 
{\it long range interactions}  and includes the exchange of gravitons and photons. With the massive particles at rest 
 they give  rise to Coulomb and Newton potentials in the non-relativistic limit. As we said, one can obtain the 
$U(1)$ WGC constraint from imposing that Coulomb repulsion dominates. In the class 2) of diagrams 
it is the massive particles which are exchanged and hence they instead are sensitive to
{\it short range interactions}. Keeping in parallel with the first class, we consider the 
massive particles almost at rest. There are three type  of tree level processes in this class, which are related by
crossing symmetry: a) Pair production of a pair of massive states (e.g. $\gamma \gamma \rightarrow \psi {\bar \psi}$)
, b) annihilation of a pair massive states  (e.g. $\psi {\bar \psi} \rightarrow \gamma \gamma$) and 3) Compton scattering.

Both classes of processes  give rise to complementary information concerning the strength of gravity versus other interactions.
In particular the second class,  which involves the propagator of a  massive state,  is  sensitive to short distance  interactions.
 Contact interactions 
exist in $d=4$ for the coupling between gauge bosons and charged scalars, $e^2A^\mu A_\mu|\phi|^2$.  However this brings no uncertainty
in the strength of the interaction, since gauge invariance relates this coupling to the trilinear gauge coupling $eA_\mu\phi^*\partial^\mu \phi$.
However, in the case of quartic scalar couplings like $\lambda |\phi|^2|H|^2$ with e.g. $\phi$ a modulus and
$H$ some massive scalar,   no information about  its strength is in general provided
by one-particle exchange diagrams. In fact such quartic interactions are known to exist in examples of BPS states of
$N=2$ supergravity \cite{Palti,Lust,DallAgata:2020ino} and hence one would like to take them into account in our
understanding of gravity as the weakest force ideas.

In order to compare the strength of some interaction induced  by some massless mediator
(gauge boson or scalar)  to gravitational interactions we propose  to use the
second class of processes involving a massive propagator. In particular we will 
consider pair production of massive states  at threshold.  The inverse process, massive particle annihilation at rest
would yield equivalent results.
 In the rest of the paper we will talk mainly about pair production but we must emphasize that all the discussion goes through 
 replacing pair production at threshold by pair annihilation at rest.
 In such  kinematical regimes both trilinear 
 and local quartic interactions (if present) are tested and may be compared with the analogous 
production mechanism from graviton production.
Strictly speaking cross sections vanish at threshold, what we will be comparing is the differential 
cross sections or rather the square of the amplitudes near threshold. 
In the case of pair annihilation
we would directly compare the annihilation cross sections at rest.

 One of the 
attractive features of this approach is that one can derive the usual WGC constraints from multiple
$U(1)$'s {\it and} a new scalar version of the WGC in a unified manner and starting from a single principle. In fact we believe that our proposal gives the first derivation of a scalar WGC from a general  underlying principle.  
Other previous discussion of a SWGC do not follow in such a direct way since in particular both scalars and gravitons 
lead  to attractive interactions 
at large distances 
and hence
no-bound-state arguments fail in this case.  One has to rely on $N=2$ SUGRA identities so that  the evidence outside the $N=2$ case becomes weaker.
Another reason to consider the production rate 
  at threshold is its possible connection with  extremal-black-hole radiation through charge pair production. 
  Or black-hole pair annihilation into photons/gravitons.
   We leave the study of this possible connection to
  future work.

  A point to note is that ours  is a quantum relativistic condition since it involves 
  particle production and interaction rates. This is unlike the case of
  one photon/graviton exchange with particles at rest which give rise to the classical non-relativistic 
  Coulomb/Newton potentials.

Specifically, the general idea may be formulated in the following terms.
Consider a theory with $U(1)$ gauge interactions or  moduli scalar fields coupled to gravity. 
Our general proposal may be stated as the

{\bf Pair Production Weak Gravity Conjecture (PPWGC):}

{\it For any rational direction in the charge lattice $\vec{Q}$ and for every point in moduli space, there is a stable or metastable particle $M$ of mass $m$  whose pair production rate by gauge or scalar mediators at threshold is larger than its graviton production rate:}
\beq
\boxed{
 |T(ij \ \longrightarrow \ MM^*)|_{\text{th}}^2 \ \geq \ |T(gg\ \longrightarrow \ MM^*)|_{\text{th}}^2}\  .
 \label{GENPPWGC}
\eeq
Here $i,j$ denotes either  $U(1)^n$  gauge bosons or scalar moduli fields and the subindex $th$ corresponds to threshold.
The criteria we propose could also  be easily generalized to theories with non-abelian gauge fields but we will not consider that possibility in the present paper.
In sections 2 and 3 in this paper the scalars will be consider massless, having in mind moduli fields. In sections 4 and 5 we will 
discuss possible extensions to the case in which the scalars are massive.

  In order to apply this principle to the case of $U(1)$ couplings we have computed the production rates 
  of charged scalars and fermions starting from photons and gravitons. 
  Production from gravitons is a non-trivial calculation. Fortunately it may be
  obtained by crossing symmetry using results for graviton Compton scattering in the
  literature \cite{holstein}. 
  The bounds obtained exactly match the results discussed for the WGC in the literature,  imposing the
  instability of extremal charged Reissner-Nordstrom black-holes. 
  We also extend the analysis to the case of multiple $U(1)$'s and argue for natural extensions, PPWGC versions of the
  Tower and Sublattice conjectures.
  
When the mediators are massless scalars one obtains  new interesting constraints.
In particular one gets a scalar WGC (SWGC) constraint  involving both trilinear and quartic scalar couplings. 
If the inequalities are saturated, one obtains 
 a differential equation involving scalar masses and their first and second derivatives. This equation is closely related to previous formulas found in \cite{Palti} and \cite{Gonzalo:2019gjp}. 
 The precise form of the SWGC conditions depends on the metric of the moduli  in the effective field theory, but some general properties of
the  extremal solutions are as follows.
\begin{enumerate}
	\item In all of the examples we study there are solutions for the massive  scalars  saturating the bounds
	which behave at large moduli like BPS-like, KK or winding  states with  built-in duality symmetries.
	This is remarkable since in the effective field theory there was no input related neither to extra dimensions, 
	extended objects nor dualities, just diagrammatics of the particles involved. These solutions are consistent with
	the Swampland Distance conjecture.
	
	\item The constraint disappears  as $M_{p}\rightarrow \infty$, unlike other versions of the WGC involving scalars
	\cite{Gonzalo:2019gjp,Freivogel:2019mtr}.
	
		\item
	The constraint is consistent with  Special Kahler Geometry identities of  $N=2$ BPS states.  We test it further in a class of Type IIA CY vacua 
in which towers of BPS particles coming from $Dp$-branes wrapping even cycles become massless at large Kahler moduli 
\cite{Grimm:2018ohb,Corvilain:2018lgw,Font:2019cxq,timo,timo2}.
They saturate our bound and feature the above mentioned duality, which in this case corresponds to T-duality.
	
\end{enumerate}

The obtained bounds apply to massive  states corresponding  to  BPS-like, KK or winding objects. 
Those are in general very heavy particles  with masses of order the Planck scale unless going to extreme limits
in moduli space. On the other hand we would like to see whether we can learn something about constraints 
on light (but not massless)  scalars which may have some relevance in particle physics or cosmology.
In this direction we briefly discuss two possibilities:

In section 4 we consider the possibility that the
potential of scalar fields (like moduli themselves) is a function of
the mass of the WGC fields, with the latter subject to the derived bounds.
In a simplified case of a single massive object one obtains interesting 
constraints having some resemblance with  the refined dS conjecture of ref.\cite{dS1,krishnan,dS3}.  
Extrema have  constraints on the second derivative 
of the potential, in agreement with the dS conjecture considerations, although  they also  apply to AdS vacua.
 In this simple one-modulus case one can show that dS minima are forbidden.

In section 5 we consider the more speculative possibility that the moduli themselves have masses 
subject to the same constraints as the WGC  states  which obey the conditions. This gives rise to constraints involving
3-d and 4-th derivatives of the scalar potential, analogous to those discussed  in ref.\cite{Gonzalo:2019gjp} but with an absolute value taken.
Some particle physics and cosmology implications from that kind of constraint were described in that reference. 
However the presence of the absolute value changes some of the consequences. In particular, the 
condition in the present case disappears when gravity decouples and no restriction on scalar field ranges appear in the 
field theory in the infrared.

The idea underlying our pair-production proposal is not to put it forward  as an alternative to long range 
one-particle exchange arguments. Our point of view is rather that the hypothesis of gravity being the weakest force could
be tested in different particle configurations and kinematic limits.  Each of them may be optimal to test a 
particular property of the WGC ideas. The Pair-Production proposal is an S-matrix criterion and is optimal to test the WGC when scalar interactions
are involved. The general idea may, in principle, be applied in any number of dimensions. Nevertheless, in this work we restrict our computations and arguments to $d=4$. The Pair-Production criteria may actually turn out to be closely related to black-hole decay and the standard WGC. Whereas 
usual WGC arguments based on stability of extremal black-holes are purely kinematical, our condition may perhaps  point 
to an additional dynamical condition.

The structure of this paper is as follows. In the next section we study the PPWGC for the case of $U(1)$ interactions. We
first compute the production rate at threshold of both charged scalar and fermion pairs from photons and gravitons.
We show how insisting on the graviton rate being smaller than the photon rate reproduces the usual $U(1)$ WGC constraint.
We also generalise the constraint to the multiple $U(1)$ case. In section 3 we study the PPWGC for scalars, and 
compute the production rate of a pair of heavy scalars from the collision of two moduli. Insisting that this rate is larger than
the rate from graviton production we obtain  the Scalar WGC constraint. We apply it to the case of complex and real scalars 
and study the structure of the massive states which saturate the bounds. 
Several examples are presented and consistency with known $N=2$ BPS results is shown.  
Section 4 study possible connections with the dS conjecture and section 5 briefly discusses 
the case of the {\it Strong} or generalized Scalar WGC's  in which the masses of the moduli are assumed to obey the same constraints 
as the massive WGC states discussed in the previous sections. Some final comments are presented in section 7.

\section{ The PPWGC for $U(1)$ interactions  }
  
\subsection{ A single $U(1)$}

In this section we study  the case of a single $U(1)$ with pair production of  charged scalars and fermions.

\begin{figure}[H]
	\centering{}
	\label{diagrama1}
	\includegraphics[scale=0.7]{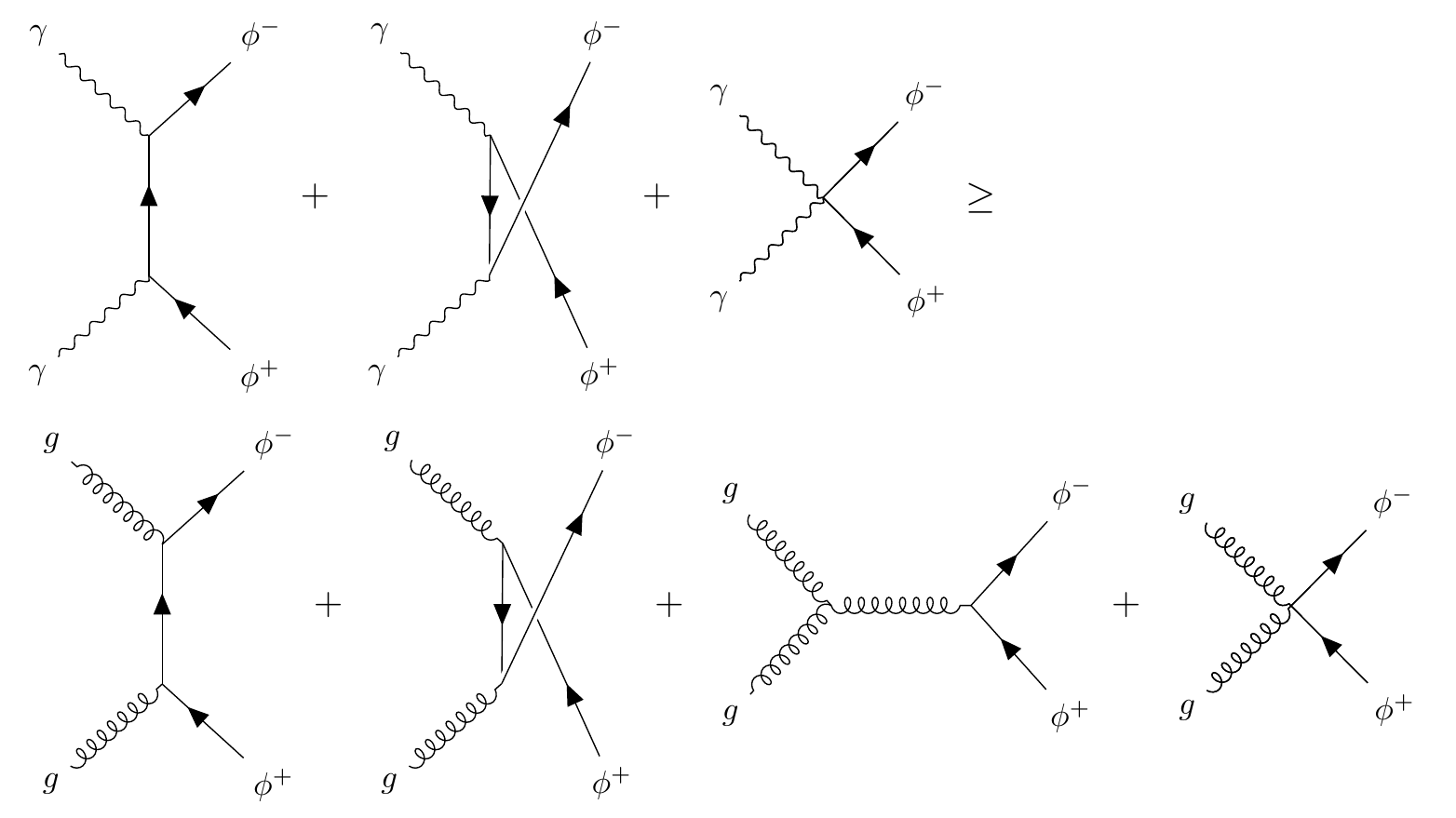} 
	\caption{\footnotesize The relevant tree level diagrams for the pair production of charged scalars in SQED and linearized Einstein gravity.  We assign the letter $A$ to the diagrams with photons and the letter C to the production via gravitons.
		}
\end{figure}

Let us start with the production of scalars.
The relevant diagrams are shown in Fig.(1).
For the photon production we are not including a diagram in which the two photons go to a graviton which then produce two scalars.
The reason is that it is purely gravitational and hence should not be included if our aim is to compare a purely electro-magnetic production 
with a purely gravitational production. Thus for the photon production we are taking the $M_p\rightarrow \infty$ limit.
The cross sections for photon and graviton pair production in the CM are written as
\beq
\left(\frac{d\sigma}{dt}\right)_{\text{CM}}^{\text{SQED}}=\frac{|A|^{2}}{32\pi s^2}\ \ ;\ \ 
\left(\frac{d\sigma}{dt}\right)_{\text{CM}}^{\text{Grav}\,\phi^{\ast}}=\frac{|C|^{2}}{32\pi s^2} \ .
\eeq
At threshold the four-momentum of the final particles is $p=\left(m,\vec{0}\right)$ and the cross section vanishes. We are not interested in comparing the cross section at threshold, but in the threshold limit, where the particles in the produced pair have infinitesimal but non-zero momenta. Thus, what we will compare is the differential cross section with respect to $t$.  In the threshold limit the Mandelstman variables are given by: $t=u=-m^{2}$ and $s=4m^{2}$. 
 Using the helicity formalism, we will see that 
at threshold only the amplitudes where the initial photons or gravitons have opposite helicities contribute. 
Both amplitudes have the structure
\beq
\left| M\right|^2\ =\ 2\left|M_{++}\right|^2\ +\ 2\left|M_{+-}\right|^2\  \ .
\eeq
For the photon production amplitude one obtains
\beq
A_{+-}=\frac{2e^{2}\left(m^{4}-ut\right)}{\left(t-m^{2}\right)\left(u-m^{2}\right)} \ \ ;\ \ 
A_{++}=-\frac{2e^{2}m^{2}s}{\left(t-m^{2}\right)\left(u-m^{2}\right)}.
\eeq
The computation of the graviton production is non-trivial.  Fortunately,  the rate may be obtained by crossing from the graviton
Compton scattering computed in \cite{holstein}. Interestingly, one finds that the gravitational amplitudes for the Compton scattering of a
spin $S$   particle with gravitons are given by the product of the electromagnetic Compton scalar amplitude times the
electromagnetic amplitude for a spin $S$ particle \cite{Choi:1993wu,Choi:1993xa,Holstein:2006pq} \footnote{ This is an avatar of the
$ (\text{gravity})=(\text{gauge})^2$ property of scattering amplitudes, see e.g. \cite{Bern:2002kj,Carrasco:2015iwa} and references therein.}:
\beq
|C_{+-}|^{2}=|C_{-+}|^{2}=F^{2}|A_{+-}|^{4} \ \ ;\ \ 
|C_{++}|^{2}=|C_{--}|^{2}=F^{2}|A_{++}|^{4},
\eeq
 where
\beq
F=\frac{1}{4M_{p}^2e^{4}}\frac{\left(t-m^{2}\right)\left(u-m^{2}\right)}{s}.
\eeq
At threshold one has $s=4m^2, t=u=-m^2$ and one obtains
\beq 
|A_{+-}|^2\ \longrightarrow \ 0, \  \, \  |A_{++}|^2\ \longrightarrow \ 4e^4 \,\,\,\, \text{and} \,\,\,\, |C_{+-}|^{2} \longrightarrow \ 0, \,    |C_{++}|^{2}\longrightarrow\frac{m^{4}}{M_{p}^4}\ \ .
\eeq
The PPWGC then gives us:
\beq
|A|^2\ \geq |C|^2 \ \longrightarrow  \ \boxed{ \sqrt{2} e\ \geq \ \frac {m}{M_p}} \ ,
\eeq
in agreement with the standard constraint of the WGC for a single $U(1)$. The factor $\sqrt{2}$ is important since it 
is precisely the factor that appears  for extremal Reissner-Nordstrom black-holes.

For completeness, 
let us consider now the spin 1/2 fermion production, although in the rest of the paper we will concentrate on the
production of scalars. The relevant diagrams are shown in Fig.(2).
%
\begin{figure}[H]
	\centering{}
	\label{diagrama2}
	\includegraphics[scale=0.7]{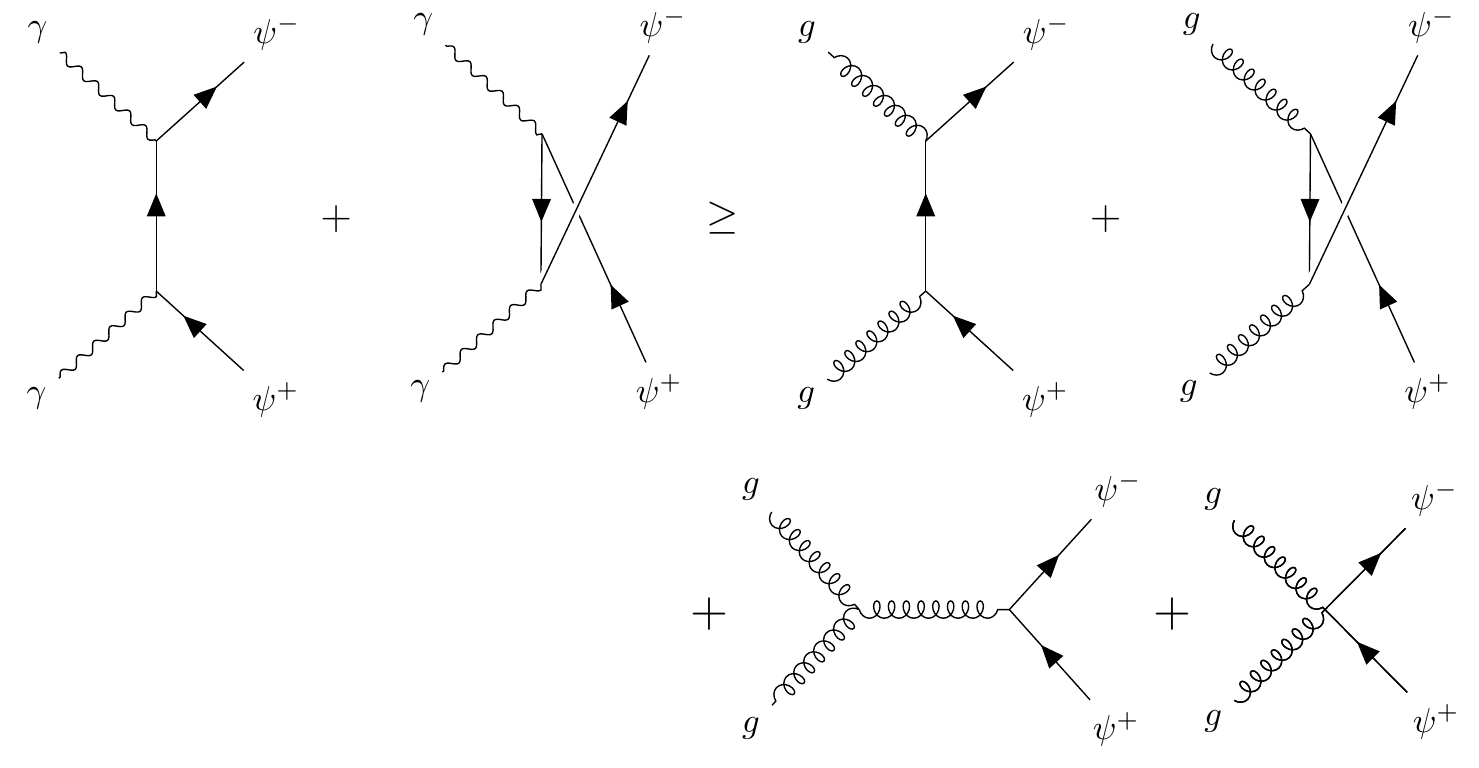} 
	\caption{\footnotesize Tree level diagrams contributing to the pair production in QED and linearized Einstein gravity. We assign the letter $B$ to the diagrams with photons and the letter D to those with gravitons.}
\end{figure}
We sum over spins in the final state in both rates. Denoting $B$ and $D$ the photon and graviton amplitudes respectively one finds
\beq
|B_{+-}|^{2}=\frac{4e^{4}\left(m^{4}-ut\right)\left[2\left(m^{4}-ut\right)+s^{2}\right]}{\left(t-m^{2}\right)^{2}\left(u-m^{2}\right)^{2}}\longrightarrow0
\eeq
\beq
|B_{++}|^{2}=\frac{4m^2 e^{4}s^{2}\left(2m^{2}-s\right)}{\left(t-m^{2}\right)^{2}\left(u-m^{2}\right)^{2}}\longrightarrow 8e^{4} \ ,
\eeq
\beq
|D_{++}|^{2}+|D_{+-}|^{2}=F^{2}\left(|A_{++}|^{2}|B_{++}|^{2}+|A_{+-}|^{2}|B_{+-}|^{2}\right)\longrightarrow\frac{2m^{4}}{M_{p}^4} \ ,
\eeq
where we have already indicated the value at threshold.
Then  PPWGC also gives us
\beq
|B|^2\ \geq |D|^2 \ \longrightarrow  \ \boxed{ \sqrt{2} e\ \geq \ \frac {m}{M_p}} \ ,
\eeq
as expected. Thus we see that,  imposing that the pair production  rate of charged particles at threshold to be larger than
the rate for the production from gravitons, we obtain the same constraint as the standard  $U(1)$ WGC. 
A proportionality between charges and masses in the rate was to 
be expected. But,  as we have shown, the fact that all precise factors match is non-trivial.
 It is also a test that the pair production at threshold of an  extremal state 
has equal probability either from photons or gravitons.
Using crossing symmetry, this also implies that the annihilation rate  of extremal  particles at rest into photons and gravitons
is the same.

\subsection {Multiple $U(1)$'s}

Consider now $N$ $U(1)$ gauge bosons with a diagonal and canonical kinetic term. We should now insist that a particle with mass $m$ and charge
vector ${\vec Q}=(Q_1,...,Q_N)$ must exist so that its pair production by photons is equal or bigger than its pair production by gravitons.
The calculation of the rates in the previous section is trivially extended for multiple U(1) and gives:
\beq
\left(Q_1^4+Q_1^2Q_2^2+Q_2^2Q_1^2+\ ..\ +Q_N^4\right)\ = \left(\vec{Q}^2\right)^2 \  \geq \ \frac {m^4}{4 M_p^4} \ .
\label{multipleU1}
 \eeq

The general statement of the PPWGC applied to this case would say that for every rational direction in the charge lattice there is a
particle of mass $m$ whose photon production rate at threshold is larger than its graviton production rate. 
Note that the produced objects must be stable or metastable particles, for the Feynman graph computation to make sense.

To shorten notation we can say that a charged state is {\bf  superproduced} if the rate to produce a pair of such particles at threshold is larger or equal to the rate to produce that pair from gravitons. Then the above conjecture may be restated as:

{\bf The Pair Production Weak Gravity Conjecture (PPWGC) for Photons.} {\it For any rational direction in the charge lattice  ${\vec Q}$  there is a (meta)stable particle
	which is superproduced}.

Here a  rational direction is a ray in the charge lattice, passing through both the origin and $\vec{Q}$.
We chose to impose the PPWGC for every rational direction in the charge lattice. A motivation for this choice is that the superproduced
 particle
 acts also as a standard WGC state to which extremal black-holes can decay. Note that the PPWGC so defined  includes the WGC but it is stronger.  
Let us review the latter to ilustrate this point. 
As stated e.g.  in \cite{repulsive} the WGC reads:

{\bf The Weak Gravity Conjecture (WGC)} {\it For every rational direction in the charge lattice there is a superextremal multiparticle state}.

A superextremal state is one whose  $\vec{Z}=\vec{Q}/m$ is either outside or on the boundary of the black-hole region. For the theory we are considering the black-hole region is simply given by $M_{\text{BH}}\leq \sqrt{2}|{\vec Q}|_{\text{BH}}M_p$.  Therefore, from (\ref{multipleU1}) we can see that, in this context with no scalar fields, superproduced is equivalent to superextremal.  Fig. 3 (an adaptation of a figure in \cite{bentalk}) illustrates the relation between the statements of the WGC and PPWGC with the well-known Convex Hull Condition (CHC). In this figure we considered a $U(1)^{2}$ with three fundamental (not composite) particles and their corresponding antiparticles in the spectra. These six particles are displayed with blue dots. The maroon dots are multi-particle states formed from them. For illustrative purposes we have written the charge and the mass of three randomly chosen states, which appear in black in the figure. The more particles a state has, the smaller the size of the dot representing it. One can see that the multi-particle states populate the convex hull of the fundamental particles in $\vec{Z}$ space. The black-hole region is represented by a grey circle in the figure. If for every rational direction there is a superextremal state, then the convex hull encloses the black-hole region.

\begin{figure}[H]
	\centering{}
	\label{fconvexhullf}
	\includegraphics[scale=0.4]{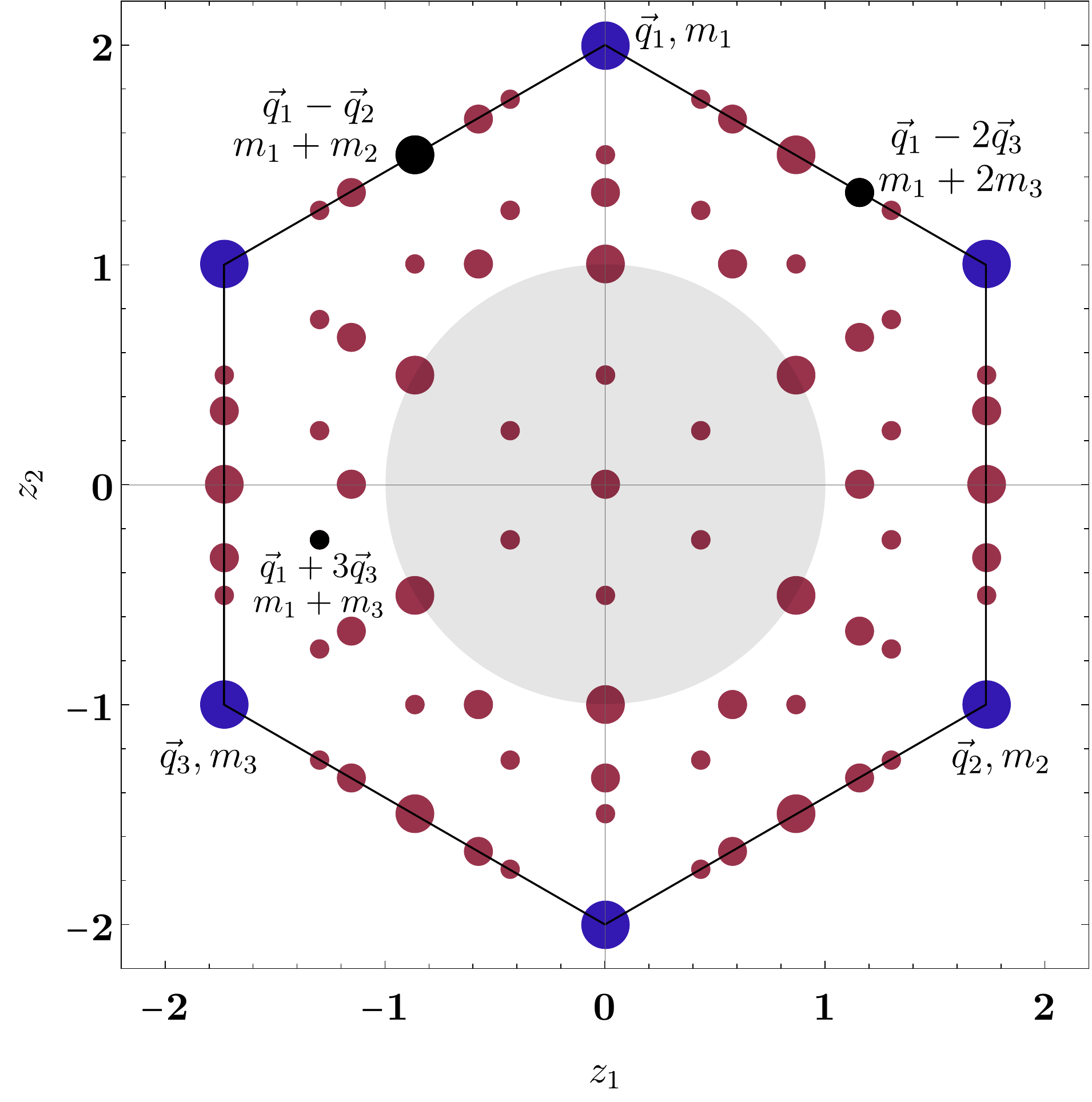} 
	\caption{\footnotesize Multi-particle states of a $U(1)^{2}$ with three fundamental particles and their corresponding antiparticles in the spectra. These six particles are displayed with blue dots. The maroon dots are multi-particle states formed from them. The more particles a state has, the smaller the size of the dot representing it. For illustrative purposes we have written the charge and the mass of three randomly chosen states, which appear in black in the figure. We use lower-case letters $\vec{z}$ and $\vec{q}$ to refer to the charges of the single-particle states. The black-hole region is represented by a grey circle. We have chosen $m_1=m_2=m_3=1$ and $\vec{q_1}=(0,2)$, $\vec{q_2}=(\sqrt{3},-1)$, $\vec{q_1}=(-\sqrt{3},-1)$. Therefore, the convex hull encloses the black-hole region in our example.
	}
\end{figure}

Notice that, unlike the PPWGC, black-hole arguments do not care whether the state is single or multi-particle. 
For us though it is not enough to have a superproduced multiparticle state to ensure that gravity is the weakest force,  we actually need a pair of particles,
possibly metastable.  Thus, the PPWGC is similar to the standard WGC, but the constraint it puts on the spectra is actually stronger than the CHC.
 The key point is that in the PPWGC approach we are producing actual particles.

It has been noted that examples from gravity and string theory suggest that a stronger version of the WGC for $U(1)^N$ is required in order to be preserved under dimensional reduction. Two closely-related strong forms are particularly well motivated: the Sublattice WGC (sLWGC) \cite{Heidenreich:2016aqi} and the Tower WGC (TWGC) \cite{Andriolo:2018lvp}. Both require the existence of an infinite number of superextremal particles along each rational direction in charge space. In this sense the PPWGC is very similar to the stronger versions of the WGC. Instead of imposing that a superproduced particle must exist for every rational direction we could have, in fact, directly imposed the tower or sublattice versions:

{\bf The Tower Pair Production WGC (TPP-WGC)}. {\it At any point $\vec{q}$ of the charge lattice there exists a positive integer n such that there is a superproduced particle of charge $n\vec{q}$}.

{\bf The Sublattice Pair Production WGC (sPP-WGC)}. {\it There exists a positive integer n such that for any site $\vec{q}$ in the charge lattice there is a superproduced particle of charged $n\vec{q}$}.

Notice that the main difference between Tower and Sublattice conjectures is that in the latter the integer $n$ is universal.
It is  interesting that the PPWGC is sensitive to whether the WGC state is a single or a multi-particle state. 

\section{Pair production from scalars  and the Scalar Weak Gravity Conjecture}

Once we have seen how the PPWGC criterium encompasses the WGC conjecture and its extensions, we will now show
how its application to production from scalars leads to interesting novel results.
 The original formulation of the WGC rested on energy and charge conservation in extremal black-hole decay. 
The absence of proper scalar charges makes a parallel reasoning difficult.
In this section we apply the principle of the Pair-Production WGC, to theories with scalar fields. 
The particular inequality which is obtained from the general formula  Eq. (\ref{GENPPWGC}) 
 will  depend on the geometry of the scalar manifold we are studying, so we will consider different possibilities.

We will take in all our examples and constraints the case of massless scalars, like moduli in string theory.
In theories with supersymmetry they may remain massless over all moduli space. So in some of the examples the massless 
scalars may be considered as a bosonic subsector of a SUSY theory. Still, the principle of gravity being the weakest form seems
unrelated to supersymmetry,  and the idea would be that the constraints obtained should also apply to non-SUSY theories in
which for some reason the scalars remain much lighter than the Planck scale.

 Let us start with the simple case of a massless complex scalar field $T$ and a complex heavy scalar field $H$ with a mass $m^2(T,T^{*})$.  The relevant part of the action has a structure
\beq
\mathcal{L_{\text{T}}}=\partial_{\mu}H\partial^{\mu}{\overline H}+ \partial_{\mu}T\partial^{\mu}{\overline T} \ -\ m^2(T,T^{*})|H|^{2} \ ,
\eeq
with a moduli dependent mass term for the heavy scalar $H$.
It is always possible to expand $m^2$ at a generic point in moduli space up to second order in the fields, and write the result in terms of either real or complex components . In terms of the complex variables:
\beq
m^2\ \simeq \ m^2_0\ +\  (\partial _Tm^2)T\ +  \  (\partial_{\overline T}m^2){\overline T}\ + \ (\partial_{\overline T} \partial_T m^2) |T|^2 +\  \frac{1}{2}\partial_{T}^{2}mT^{2}+\frac{1}{2}\partial_{\bar{T}}^{2}m\bar{T}^{2}...
\label{desarrollo}
\eeq

\begin{figure}[H]
	\centering{}
	\label{diagrama3}
	\includegraphics[scale=0.7]{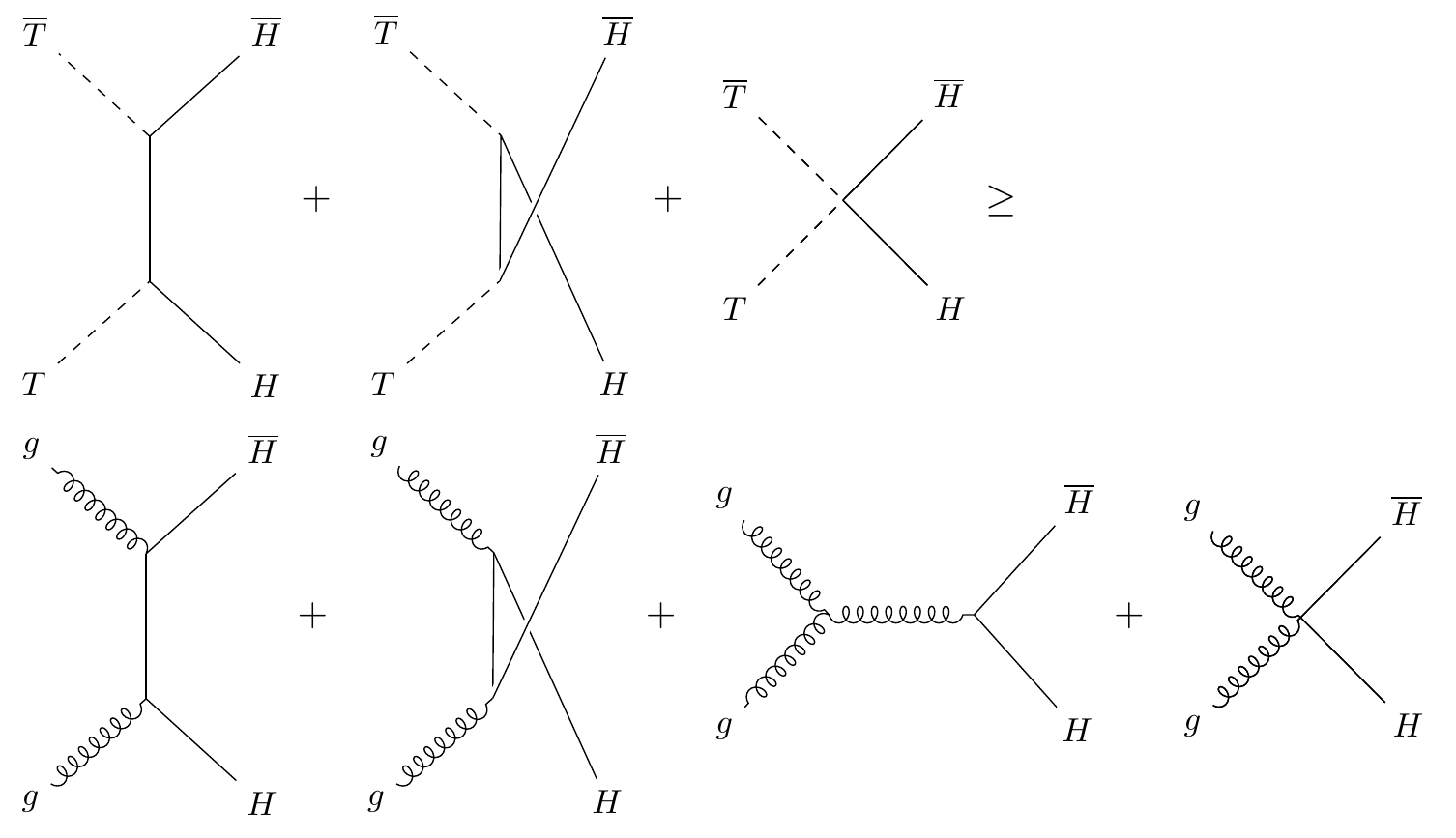} 
	\caption{\footnotesize Tree level diagrams contributing to the pair production in the scalar theory and linearized Einstein gravity. We assign the letter $N$ to the diagrams with moduli. }
	\label{diagscalar}
\end{figure}

Following the PPWGC, we ought to consider the pair production of the field $H$ and compare it with the production from gravitons at threshold. The relevant diagrams  for the process  $T{\overline T}\rightarrow H{\overline H}$  are shown in Fig.(\ref{diagscalar}).
Again, in the production from scalars we are not including the annihilation of two scalars into a graviton, with the latter producing 
heavy pairs, since it  is purely gravitational and does not involve scalar couplings. Notice also that the last two terms in the $m^2$ expansion will not contribute to the four point function that we are interested in, where the initial particles are a pair $T$, ${\overline T}$. From the expansion we extract the trilinear  $\Delta T|H|^2  +h.c.$ and the quartic $\lambda |T|^2|H|^2$ couplings:
\beq
\Delta  = {\partial_T m^2} \ , \  {\overline \Delta } = {\partial_{\overline T}m^2} \ \ ,\ \ \lambda  =  \partial_T\partial_{\overline T}m^2 \  .
\eeq
The amplitude has the form:
\beq
N\ =\  -|\Delta|^2\left[\frac {1}{t-m^2}+\frac {1}{u-m^2}\right]\ -\ \lambda
\eeq
The gravitational diagrams are the same as in Section 2. At threshold one has $t=u=-m^2$, and the condition reads
\beq
 \left| \frac {|\Delta|^2}{m^2}\ -\  \lambda \right| ^2 \ \geq \ \frac {m^4}{M_p^4} .
 \label{scalarppwgc}
\eeq
In terms of mass derivatives one obtains 
\beq
\left| (\partial_T m^2)(\partial_{\overline T}m^2) \ -\  m^2 \ \partial_T\partial_{\overline T}m^2\right| \  \geq \ \frac {m^4}{M_p^2} \ .
\label{unocomplex}
\eeq
For $n$ complex moduli $T_i$, $i-1,..,n$ parameterising a hermitian manifold with a metric $g_{i{\bar j}}$ this is generalised to
\beq
\boxed {  \frac { g^{i{\bar j}} }{n}  \ \left| (\partial_i m^2)(\partial_{\bar j}m^2)\ -\ m^2 (\partial_i\partial_{\bar j} m^2)\right|\ \geq \  \frac {m^4}{M_p^2}   }  \ .
\label{varioscomplex}
\eeq
This is the general form of the scalar WGC for n complex moduli.
Notice that, as expected,  this expression is  invariant under holomorphic coordinate transformations. We could replace the partial derivatives with covariant derivatives, however, nothing would change since the mixed index components of the connection vanish in a hermitian manifold.  
In order to compare the contribution of the moduli to graviton production an averaging factor $1/n$ is included. In other words, the contribution of all moduli
should be compared with  $n$-times the production rate from gravitons in order to have a fair comparison.
Such an averaging was not needed in the case of production from photons in an $U(1)^n$ theory with canonical kinetic basis because for any 
given charged particle one can always find a basis in which it couples to a single $U(1)$.
 Note  that '$n$' here refers only to {\it active} moduli i.e. the subset
of the moduli in the theory which couple to a particular massive state.
A graphical interpretation of this constraint is given in Fig. (5). The region between the two parallel lines is forbidden, but it disappears 
as $M_p\rightarrow \infty$, as is expected from a swampland condition. Points in field space saturating the bound lie on top
 of the blue lines marking the boundaries.
\begin{figure}[H]
	\centering{}
	\label{barra1}
	\includegraphics[scale=0.3]{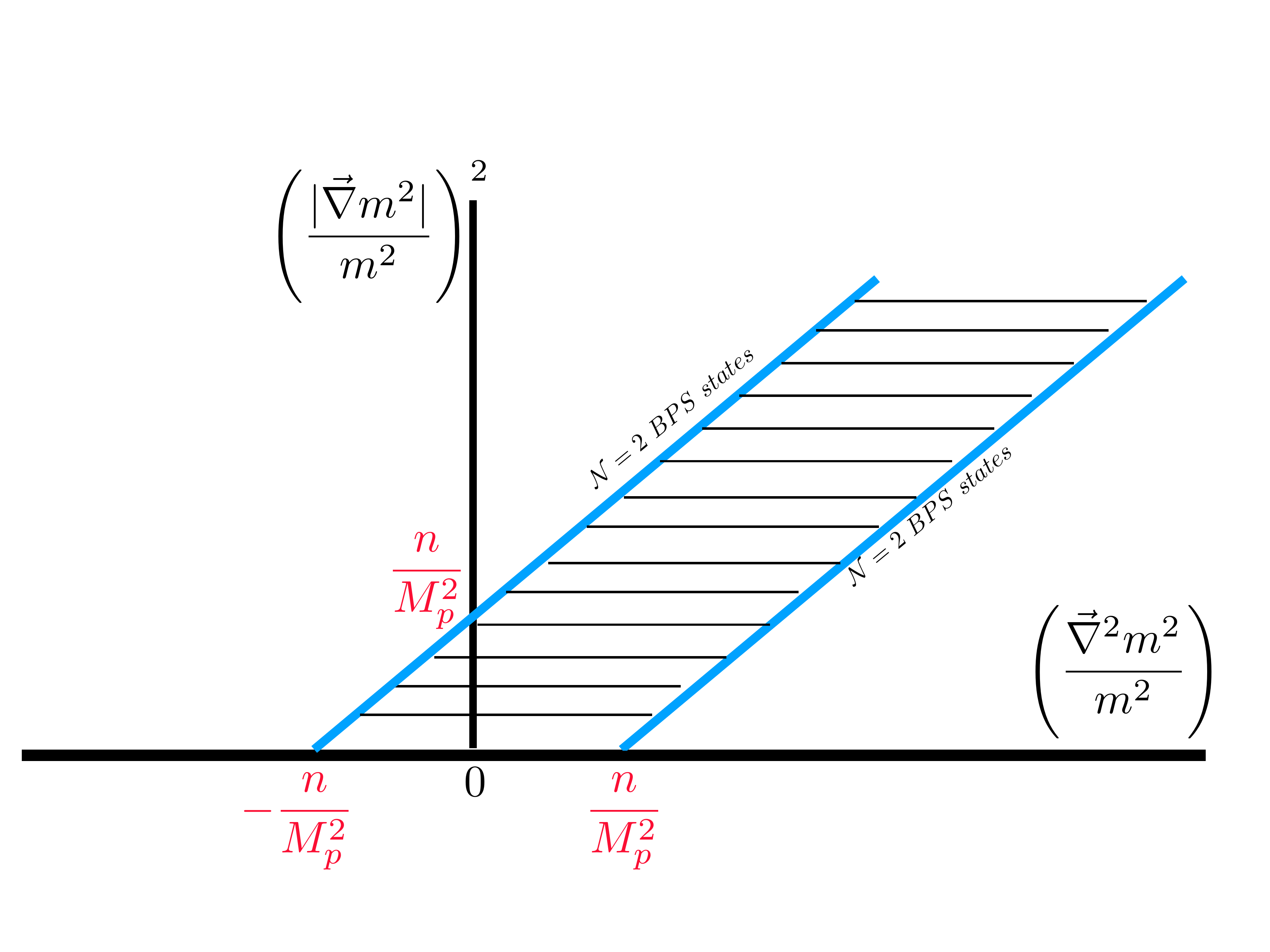} 
	\caption{\footnotesize The scalar WGC for $n$  active moduli  coupling to  massive WGC scalars. The barred region correspond to 
	points in moduli space in which gravity is too strong. This region disappears as $M_{p}\rightarrow \infty$. $N=2$ BPS states 
	lie on top of the blue  boundary lines. 
	}
\end{figure}

One can then state 

{\bf The Pair Production Scalar Weak Gravity Conjecture (PPSWGC).} {\it  Given  any set of moduli scalars,  there must be a massive  
particle $H$ with mass $m$  coupled to them such that their average
production rate at threshold from moduli is larger than the corresponding rate from gravitons.}


\subsection{Examples}

\subsubsection{ Complex scalar in $N=1$ supergravity}

Consider first the case of $N=1$ supergravity with a metric $g_{i{\bar j}}=K_{i{\bar j}}$, with $K(T,T^*)$ the Kahler potential.  Without loss of generality 
let us define the real function $F(T_i,T_i^*)$ by $m^2=M_p^2e^F$, and take n complex $T_i$ fields dimensionless.
One can check  that Eq. (\ref{varioscomplex}) may be rewritten in the
simple form
\beq
\boxed{ g^{i{\bar j}}\left|F_{i{\bar j}}\right| \ \geq \ n }\ .
\label{seminal}
\eeq
\\
Note that, due to the absolute value, there is a symmetry under $F\leftrightarrow -F$. This tells us that, if there is a particle with mass $m$
verifying the bound, a particle with mass $m'=M_p^2/m$ would also obey it. In the specific
models below this symmetry would correspond to a duality symmetry. Note also  that in a $N=1$ supergravity 
theory with spontaneously broken SUSY,  the 
gravitino mass may be written as $m_{3/2}^2=e^G$, with $G$ the full Kahler potential. With this structure such a mass automatically
saturates the  bound, which would apply rather to the scalar s-Goldstino, since the  massive states in our derivation are scalars.

Let us consider the simple case in which the moduli have a no-scale metric,
i.e., $g_{i,{\bar j}}=\delta_{i,{\bar j}}/(T_i+T_i^*)^2$.
These appear
for example  in $N=1$  toroidal/orbifold compactifications down to 4D in string theory (see e.g.\cite{BOOK}). The conjecture requires now the existence of 
scalar fields, with mass $m^2(T_i,T_i^*)$, coupled to the moduli. The constraint is in this example 
\beq
\delta ^{i{\bar j}} (T_i+T_i^*)^2\ \left| F_{i{\bar j}}\right|\ \geq \ n \ .
\eeq
Let us study the case in which the inequality saturates. One finds solutions
\beq
F\ =\ f(T_i)+f^*(T^*_{\bar i})  \ + \  \sum_i  \eta_i \text{log}(T_i+T_i^*)
\eeq
with $f(T_i)$ an arbitrary holomorphic function. Here $\eta_i$ takes all possible choices $\eta_i =\pm 1$. 
In this case our PPWGC saturating  states would have a mass
\beq
m^2_a \ =\  m_0^2|e^{f(T_i)}|^2 t_1^{\pm 1}t_2^{\pm 1}...t_{n}^{\pm 1} \ .
\label{tpowers}
\eeq
with $f(T_i)$ an arbitrary holomorphic function and $t_i=2 \, \text{Re}T_i$. The index $a=1,2,..2^{n}$.  
The large modulus behaviour depends on the form of the holomorphic functions $f(T_i)$.
In the case of constant $f$'s,
going to a canonical frame with $t=e^\sigma$ there are states which become exponentially light 
in the limits $\sigma \rightarrow \pm \infty$. This behaviour would be in agreement with the expectations of the 
swampland distance conjecture. Also for each saturating state there is another dual state with inverse mass, as
pointed out above. We will see below that certain classes of BPS states in known $N=2$ supergravity theories from string theory
are consistent with this structure. 

For more general CY the metric of the Kahler moduli (in. e.g. Type IIA string theory) has the behaviour at large moduli
\beq
K_{i{\bar j}}\ \simeq \ \frac {d_i}{(T_i+T_{\bar i})^2} \ ,
\eeq
where the $d_i$ are integers characteristic of each singular limit \cite{Grimm:2018ohb,Corvilain:2018lgw,Gendler:2020dfp}.
From Eq. (\ref{seminal}) one can compute the asymptotic behaviour of the particles which saturate our bound. One now finds
\beq
m^2_a \ =\  m_0^2|e^{f(T_i)}|^2 (t_1^{\pm d_1}...t_{n}^{\pm d_n}) \ .
\eeq
This behaviour,  corresponds e.g. to the asymptotic behaviour found in  \cite{Grimm:2018ohb,Gendler:2020dfp}, showing 
the large moduli regime  of BPS states in Type IIB CY compactifications.
It would be interesting to go through examples in e.g. \cite{Grimm:2018ohb,Corvilain:2018lgw,timo,Font:2019cxq,timo2} 
and check the agreement with the constraint.

The solutions in (\ref{tpowers}) allow for alternative behavior  depending on the
particular holomorphic functions $f(T_i)$, the arbitrariness is substantial.
For example,  one may chose all $\eta_i=-1$ and 
$f=-\text{log}(\Pi_i\eta(T_i)^2)$, with $\eta$ the Dedekind function. With such a choice the mass is $SL(2,{\bf Z})^n$ invariant.
This kind of structure appears in the class of duality invariant non-perturbative potentials considered in \cite{Gonzalo:2018guu}
and references therein.  At large moduli the
Dedekind function has an exponential behaviour $\eta \sim e^{-(\pi/12)t}$, so that there could be saturating
solutions with a behaviour $m^2\sim 1/(\Pi_it_i\ e^{-\pi/3 t_i})$, exponentially growing at large $t_i$. 
This class of solutions
would not have the asymptotic behaviour of the distance conjecture, as explained in \cite{Gonzalo:2018guu}. 
Note in this respect that such exponential of exponential  behaviour at large moduli
appears also for   the states  called of Type II  in  \cite{Grimm:2018ohb,Corvilain:2018lgw} for Type II CY compactifications.

\subsubsection{Examples from BPS states in $N=2$ supergravity}

We would like now to show that examples of BPS states in $N=2$ supergravity theories from string theory saturate our bound. 
We will consider for illustration a class of BPS states which appear in Type IIA CY compactifications from $Dp$-branes wrapping even
cycles. These (and their IIB mirrors)  have been discussed in
\cite{Grimm:2018ohb,Corvilain:2018lgw,Font:2019cxq,timo},\cite{Gendler:2020dfp} to provide string theory tests of the swampland distance conjecture.
We follow here \cite{Font:2019cxq}.   The relevant   masses are summarized in the table. 
\begin{table}[h!!]\begin{center}
\renewcommand{\arraystretch}{2.00}
\begin{tabular}{|c|c|c|c|c|}
\hline
 & D0  &   D2 & D4  & D6  \\
\hline \hline
Mass$^2$ (CY) &  $e^{K_K}$   &  $e^{K_K}\left| T_a\right|^2$ &   $e^{K_k} \left| \frac {1}{2} \sum_{b,c}\kappa_{abc}T^bT^c\right|^2$ & 
$e^{K_k} \left| \frac {1}{6} \sum_{a,b,c}\kappa_{abc}T^aT^bT^c\right|^2$  \\
\hline
 Mass$^2$ ($Z_2\times Z_2$) &  $\frac {1}{t^1t^2t^3}$. & $\frac {t^i}{t^jt^k}$ & $\frac {t^jt^k}{t^i}$  & $t^1t^2t^3$ \\
 \hline
\end{tabular}
\caption{Masses of the different particles obtained by wrapping one kind of  Type IIA $\text{D}p$-brane around a given even cycle on 
a general Calabi-Yau threefold
and for the $Z_2\times Z_2$ orbifold example from ref.\cite{Font:2019cxq}. Masses are in units of $8\pi M_p^2$.}
\label{tablaBPS}
\end{center}
\end{table}  
Here $K_K$ is the Kahler potential of the Kahler moduli $T_a=t_a+i\eta_a,a=1,..,h_{11}$,
and $\kappa_{abc}$ are the triple intersection numbers in the CY.  
The masses of the BPS states may be written as $m_r^2={8\pi}e^{G_r}$, where
\beq
G_r\ =\ \text{log}|W_r|^2 \ + \   K_K\ ,
\eeq
and $|W_r|^2$ is given by the different superpotential factors in  the table.  With this form one obtains
the constraint is
\beq 
g^{i{\bar j}}\ \left|(G_r)_{i{\bar j}}\right|=\ h_{11} \ ,
\eeq
which holds, since $(G_r)_{i{\bar j}}=(K_K)_{i{\bar j}}=g_{i{\bar j}}$. More explicitly, for the case of the $Z_2\times Z_2$ toroidal example considered
in \cite{Font:2019cxq} there are  three Kahler moduli  $T_i$ and 8 BPS states corresponding to D0,D2,D4 and D6 wrapping even cycles.
Their masses are $m_r^2={{8\pi}}(t_1^{\pm 1}t_2^{\pm 1}t_3^{\pm 1}$), with $t_i=2 \,\text{Re}T_i$, as shown in table \ref{tablaBPS}. 
Note that these masses agree with the result we showed in Eq. (\ref{tpowers}) (for $f$ constant) which do saturate our bound.

Note that, in agreement with the duality symmetry $F\leftrightarrow -F$, for each BPS example in the table with mass $m$, there is another one
with mass $1/m$. From the D-brane perspective, a duality with respect to the six compact dimensions transforms $D_0\leftrightarrow D_6$ and
$D_2\leftrightarrow D_4$. This is also an electric-magnetic duality since the states have also inverse charge under  magnetic $U(1)$'s.
In Fig. (5) particles coming from $D0,D2$ are points within the rightmost blue line, whereas  those coming from 
$D4,D6$ are inside the blue line on the left.

The fact that for these $N=2$ BPS states our condition is saturated is not surprising due to the following fact.
In $N=2$ supergravity the central charge $Z$  satisfies the algebraic equation
\cite{Ceresole:1995jg,Palti,DallAgata:2020ino}
\beq
g^{i{\bar j}}\left( D_i{\overline D}_{\bar j}\ |Z|^2 \ -\ D_i Z{\overline D}_{\bar j}{\overline Z}\right)\ =\ n_V|Z|^2 
 \label{SKG}
 \eeq
where $n_V$ counts the number of vector multiplets. 
This condition should be verified by the central charge of  any BPS state.
%
The above algebraic expression may be derived from the Special Kahler Geometry identities
(see e.g.\cite{Bellucci})
\beq
D_{\bar i}Z\ =\ 0 \ \ ;\ \ D_i {\overline D}_{\bar j}Z\ =\ g_{i{\bar j}}Z \ .
\label{SKG2}
\eeq
Identifying $Z$ with the ADM mass $m$  suggests to write 
\beq
m^2\ g^{i{\bar j}}\partial_i \partial_{\bar j} m^2 \ -\ (\partial_i m^2)(\partial^i m^2)\ =\ n_Vm^4 \ .
\eeq
This equation is consistent with our equation (\ref{varioscomplex}) above and masses saturating it 
would lie at the boundary blue lines in Fig. (\ref{barra1}).

\subsubsection{The case of n real scalar fields}

Let us consider now the case of  $n$ real scalar fields $t_i$ with diagonal kinetic terms.
One can obtain the trilinear and quartic couplings from the general expansion
\beq
m^2(t_i) \ \simeq \ m_0^2 \ + (\partial_i m^2)\ t_i +\ \frac {1}{2} ( \partial_i ^2 m^2)\  t_i^2\ +\ ...
\label{desarrollito}
\eeq 
Consider first   the case of  $n$ real scalars with  diagonal no-scale kinetic metrics  $g_{ii}=1/t_i^2$.
From the pair production constraint we now obtain 
\beq
\sum_i  g^{ii}\left| (\partial_im^2)^2 \ -  m^2( \partial_i^2m^2)\right| \ \geq \ n\  \frac {m^4}{M_p^2} \ .
\label{vieja}
\eeq
Writting $m^2=e^F$, when the inequality is saturated one obtains
\beq
 \sum_i (t_i^2)\  \left| \partial_i^2F\right|  \ = \ n \ ,
\label{nreales}
\eeq
with solutions
\beq
m^2(t_i)\ =\ m_0^2\left(t_i^{\pm 1}...t_n^{\pm 1}\right)e^{\sum_ib_it_i} \ .
\eeq
Here $m_0$ and $b_i$ are real integration constants.
Note that now, unlike the SUSY case above, the existence of a scalar moduli space is not expected,
and the interpretation of the massive states saturating the bounds is not obvious.
Still, it is interesting to explore for comparison what is the form of the saturating masses in this case.
We have two classes of solutions. For  $b_i=0$  one obtains saturating  states 
very similar to the complex no-scale metric example in Eq. (\ref{tpowers}),
which is the kind of behavior of BPS, KK and winding states in string theory. 
As in the previous examples, 
this is in itself remarkable, since 
it means that the  scalar PPWGC condition for  massless particles with  scale invariant metrics must come along with a massive 
spectrum behaving like winding and momenta, i.e., string theory. 
Note also that here the presence of both winding and momenta (and hence duality)  is a consequence of the invariance under $F\rightarrow -F$  of the
rates. Thus having rates and no amplitudes in comparing the interactions is at the root of the built-in duality of the massive spectrum.
Going to a
canonical frame with $t=e^{\sigma }$ the behavior at large $\sigma$ is exponential, consistent with the distance swampland conjecture.

For $b_i\not=0$ there are additional saturating  solutions.
They have an additional exponential factor  in $t$, which means exponential of exponentials once one goes to a canonical frame.
This is analogous to the result above for complex moduli.

We may alternatively consider a canonical metric for the scalar fields.
It is easy to see that in this case one obtains saturating solutions of the form
\beq
  m^2(t_i)\ =\ m_0^2\ ( t_1^{d(t_1)^{\pm}}...t_n^{d(t_n)^{\pm}}) \ \ ,\ \ d(t_i)^{\pm}\equiv c_{\pm}^i\pm  1/2log(t) \ ,
\label{single}
\eeq
where we have defined  $t_i=e^{\phi_i}$, with $\phi$ the canonical fields 
and  $c_{\pm}^i$ and $m_0^2$ are real constants.
The structure  has also a   form proportional to powers $t^{d(t)}$,  reminiscent of the
of the examples discussed above, but now with a slowly varying exponent $d(t)$. 
Under a duality transformation  $t_i\rightarrow 1/t_i$ one obtains a new solution exchanging 
$c_{+}^i \rightarrow -c_+^i$, $d(t_i)_+\rightarrow -d(t_i)_+$. So again for each solution of mass $m$
there is another solution with mass $1/m$. Note however that going to a canonical frame shows that 
the solutions have a Gaussian behavior. In fact such Gaussian solutions would also appear for 
the case of complex scalar fields with a canonical, instead of no-scale metrics. 

\subsubsection{Previous formulations of the SWGC}

There have been previous formulations of scalar weak gravity conjectures in the literature. Palti was the first in making the proposal \cite{Palti}  that a theory
with moduli $t^i$ should have a state  $H$ with mass $m$ obeying the bound
\beq
g^{ij}\ (\partial_{t^i}m)(\partial_{t^j}m)\ > \ \frac {m^2}{M_p^2} \ .
\label{minimalpalti}
\eeq
This has the simple interpretation of imposing that a trilinear $t^i|H|^2$ coupling squared is stronger than the gravitational coupling. In that paper it was noted that this inequality cannot be directly deduced from bound states arguments (or from the RFC) since both the scalars and gravity act attractively.
It was also noticed that, at large field, this expression is consistent with the swampland distance conjecture. 
Palti also proposed  (see footnote in \cite{Palti}) the inequality
 \beq
 \frac {1}{2}g^{ij}\nabla_{t^i}\nabla_{t^j}m^2 \ -\ g^{ij}(\partial_{t^i}m)(\partial_{t^j}m)\ \geq \ n\ \frac {m^2}{M_p^2}\ ,
 \label{paltiN2}
 \eeq
 with $n$ the number of real scalars coupling to the WGC state of mass $m$. The motivation for this  inequality mainly came from 
 the Special Geometry identities in $N=2 $ supergravity mentioned above. 
 Note that it is analogous to our constraint  except for the fact that we have an additional absolute value taken
 in the left  (and there is some numerical factor). 
  
In ref.\cite{Gonzalo:2019gjp}  it was proposed a slightly different version of a scalar WGC for a real scalar
with canonical metric given by
\beq
 2 (\partial_\phi m^2)^2 \ -\ m^2  \partial_\phi^2 m^2 \ \geq \  \frac {m^4}{M_p^2}\  .
 \label{viejaseminal}
\eeq
The motivation was to modify the original  scalar WGC of Eq. (\ref{minimalpalti}) to include quartic scalar
interactions. A further motivation was the intriguing structure of its saturating solutions. Indeed the above equation may be rephrased 
as 
\beq
 \rho''\ =\  \frac {\rho }{M_p^2} \ ,\ \rho\equiv \frac {1}{m^2} \ ,
 \eeq
and the saturating solutions have the form
\beq
m^2\ =\ \frac {1}{ae^\phi\ +\ be^{-\phi} }\ .
\label{extremalviejo}
\eeq
At both limits $\phi \rightarrow \pm \infty$ the behaviour is consistent with the swampland distance conjecture. 
Furthermore, defining $t=e^\phi$,  there is built-in duality under  the exchange $t\leftrightarrow 1/t$. 
Thus the saturating solutions have the structure of KK  and winding momenta, implying the existence of an underlying theory
with extended objects. This is in fact the kind of structure that we have found in the present article, although the 
precise form of the constraint is not the same. 

For a single real massless modulus and a massive state of mass $m$, the scalar PPWGC gives rather the constraint Eq. (\ref{vieja}) 
\beq
\left| (\partial_\phi m^2)^2 \ -  m^2( \partial_\phi^2m^2) \right| \ \geq \  \frac {m^4}{M_p^2} \ .
\label{viejauno}
\eeq
Comparing with (\ref{viejaseminal}) we see  there is a factor 2 missing in the first term and 
the absolute value taken on the left. It is the factor 2 which makes the solutions in (\ref{extremalviejo}) 
different from those of Eq. (\ref{single}) taken for a single field. It is an interesting question 
whether a scalar theory exists yielding a result analogous to (\ref{viejaseminal}) from
scattering amplitude arguments. 
%
%
%

   In part motivated by \cite{Gonzalo:2019gjp}, there have been some attempts to arrive at a SWGC using bound states arguments by somehow introducing short-range repulsive scalar interactions. In order to compare short with long range forces, one needs to fix an energy scale. In \cite{Benakli:2020pkm} a modified version of the RFC was proposed where only the leading interaction is to be compared with gravity. In this way they were  able to motivate differential inequalities for the SWGC.
   A different proposal was made in \cite{Freivogel:2019mtr}. They argued against the formation of gravitationally bound states with sizes smaller than their Compton wavelength. This idea was coined as Bound State Conjecture. The latter does not give rise to a differential inequality and it remains non-trivial even when gravity is turned off.   
 
The fact that the  PPWGC gives a well defined rationale for the existence of a Scalar Weak Gravity Conjecture is an important result of the present work. So far, it is the only criteria that is translated into a differential constraint including both first and second derivatives of the mass, making  direct contact with
known $N=2$ BPS constraints, proposing also a generalization   to non-SUSY settings.

 \section{Constraints on the scalar potential for moduli}
 
 In the above we have seen how the PPWGC applied to scalars suggests the existence of massive scalars which 
obey or  saturate the bounds,  so that gravity is the weakest force. These fields correspond to scalars belonging to BPS multiplets when there is
 enough SUSY.  However,  we would like to know whether any constraint may be obtained for other scalars like
 moduli themselves, once they get a mass. In particular, it would be interesting to see whether the above bounds may give
 us some constraint on moduli  (or other scalars) effective potentials.

 One possible connection, inspired by our experience in string theory, is as follows.
 Moduli  $t_i$ in string theory are massless classically and get a potential at the quantum level. Such  potentials  often appear after
 summing over loop contributions of massive $H_a$  particles, like e.g., towers of BPS states. The dependence on massive BPS states 
 may also appear at the non-perturbative level. Those massive particles have masses $m_a(t_i)$ which are functions of the moduli already
 at the classical level, we saw  some $N=2$ examples above. In those cases the induced moduli potential  will depend on the 
 moduli through the masses of the heavy BPS-like states, $V=V(m_a(t_i))$. If we insist that the masses of  heavy $H_a$ scalars $m_a$ are
 subject to the PPSWGC, one might hope to obtain some constraint on the form of the resulting  moduli 
 potential. In string theory we typically have plenty of moduli and infinite towers of BPS objects so the task is not easy. 
 Here for simplicity we are going to consider the,  admittedly,  oversimplified case of a single modulus whose potential is a function
 of a single massive state $H$ whose mass $m$ obeys a single field  version of the constraint Eq. (\ref{varioscomplex}).
 
 In this section we examine whether the scalar WGC leads to constraints similar to those of the dS conjecture.
 One possible hist of this connection is the appearance in both of the second derivative of the potential.
 Let us first recall the swampland dS conjecture  \cite{dS1,krishnan,dS3} for later comparison.
 The latter states that the  scalar  potential  for a theory coupled to gravity
satisfies either 
\beq
|\nabla V|\ \geq c\frac {V}{M_p} \ , \ \text{or} \ , \text{min}(\nabla_i\nabla_jV) \ \leq \ -c'\frac {V}{M_p^2} \ .
\label{refined}
\eeq
Here $c,c'$ are constants of order one. In the second alternative one has the minimum eigenvalue of the Hessian in an orthonormal frame.
This refined dS conjecture has the property that dS maxima are allowed (as it should since e.g. the SM has one such maximum) but dS minima are not.
This form of the dS conjecture is motivated by arguments which use the covariant entropy bound applied to a dS configuration, see \cite{dS1,krishnan,dS3}.
 
 Let us consider for definiteness the case of a $N=1$ supergravity theory with a single modulus $T$. It gets a potential at the 
 quantum level from a massive state  with mass $m^2(T,T^*)$, so that the modulus potential depends on the moduli only through 
 its dependence on this mass,  $V=V(m^2(T,T^*))$.  To simplify notation define $y=m^2$. 
  Then it is easy to check that
 \beq
 m^2_T=\frac {V_T}{V_y} \ ,\ m^2_{\overline T}=\frac {V_{\overline T}}{V_y} \ ;\ 
 m^2_{T{\overline T}}= \frac {V_{T{\overline T}  }   }   {V_y}\ -\ 
 \frac {V_{yy}}{V_y}\left(\frac {V_TV_{\overline T}}{V_y^2}\right)
 \eeq
 Imposing the PPWGC bound in Eq. (\ref{varioscomplex}), one gets the result
 \beq
 g^{T{\overline T}}\left| \frac {V_TV_{\overline T} } {V_y^2}\ (1\ +\ y\frac {V_{yy}}{V_y})\ -\  y\frac { V_{T{\overline T}  }  }{V_y} \right|
 \ \geq \frac {y^2}{M_p^2} \ .
 \eeq
 If we search for extrema $V_T=V_{\overline T}=0$  one obtains (assuming $V_y\not=0$)
 \beq 
 g^{T{\overline T}}\left|V_{T{\overline T }}\right| \ \geq\  \frac {\left| yV_y\right|}{M_p^2} \ .
  \eeq
 One sees that the second derivative of the potential is bounded below. This is reminiscent of the dS conjecture refinement,
 that if there is a extremum, the second derivative of the potential must be large enough. However, in the
 present case it applies both to dS and AdS.
   
\begin{figure}[H]
	\centering{}
	\label{barra2}
	\includegraphics[scale=0.3]{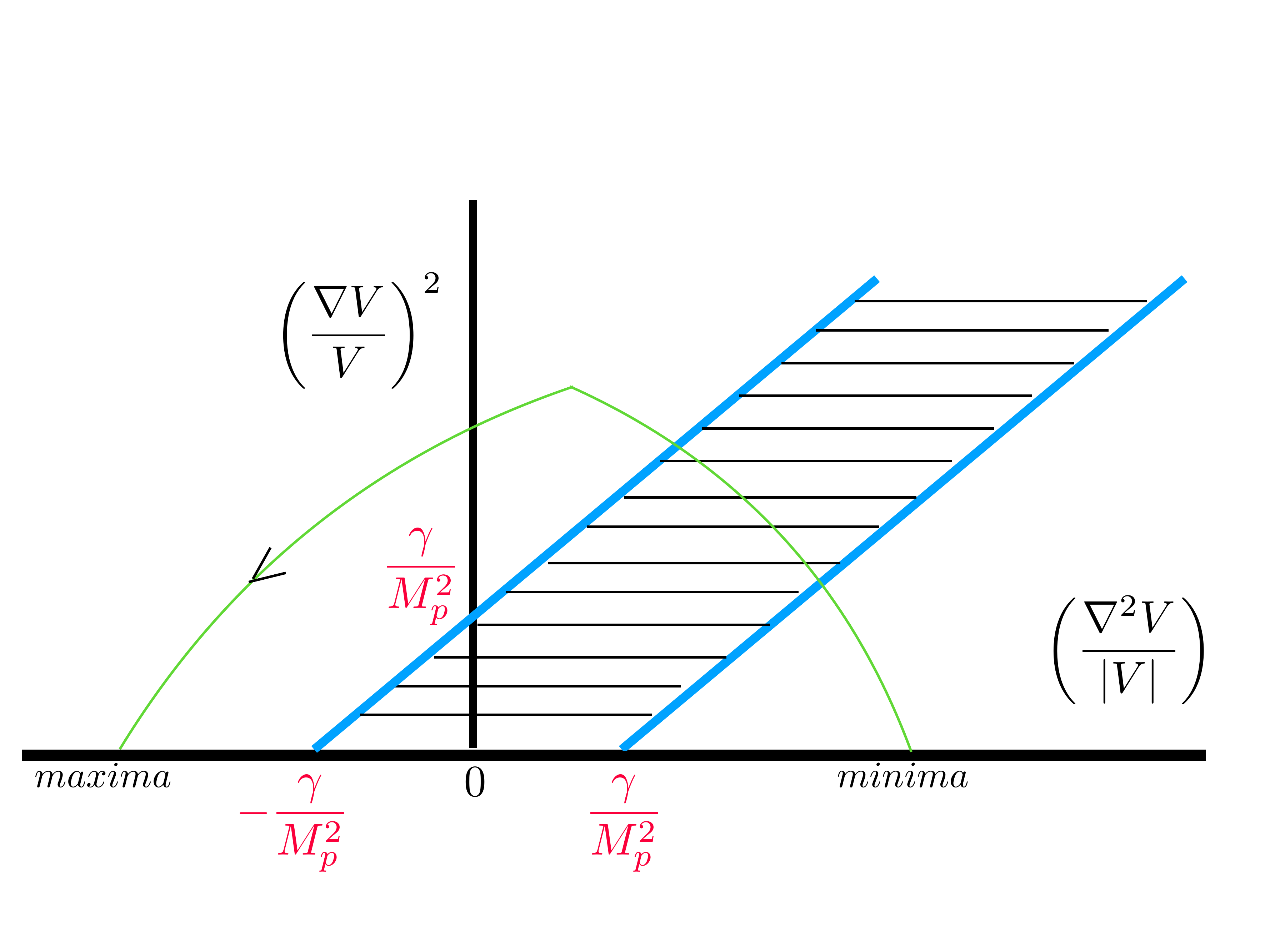} 
	\caption{\footnotesize Constraints for the potential of a single modulus when $V\propto m^{2\gamma}$. In the barred region 
	the gravitational interaction is stronger than the scalar one.  If a dS minimum exists, and the potential
	vanishes at infinity, there must be also a local maximum. This is not possible because 
	going from a minimum to a maximum in field space (green line) 
	one has to go through the forbidden region. The constraint disappears as $M_{p}\rightarrow \infty$. 
	}
\end{figure}

 More specific results are obtained if one assumes a power dependence for the potential,i.e., $V=\eta m^{2\gamma}$, with $\gamma$ a positive number
 and $\eta =\pm 1$ (see also \cite{CERN,Andriot:2020lea}). Examples of Type IIA  orientifolds with fluxes \cite{DeWolfe:2005uu,Camara:2005dc,Aldazabal:2006up} scale like 
 $V\sim m^2$ at the minima. This behavior is also a prediction of the AdS conjecture in ref.\cite{Lust:2019zwm}, recently tested in
 e.g.  \cite{Blumenhagen:2019vgj,Font:2019uva,Junghans:2020acz,Buratti:2020kda,Marchesano:2020qvg} within string theory.
 Another  example of this kind of dependence is the case of the Coleman-Weinberg one-loop potential, which is 
 proportional to the 4-th power of the mass propagating in the loop. 
 For $V\sim m^{2\gamma}$ one finds
 \beq
  \left| \left( \frac {\nabla V} {V} \right)^2\ -\  g^{T{\overline T}}\ \frac {V_{T{\overline T}}}{|V|}\right|\ \geq \ \frac {\gamma}{M_p^2} \ .
 \eeq
 At extrema one gets the condition 
 \beq
  \left|  g^{T{\overline T}}\  {V_{T{\overline T}}}\right|\ \geq \ \gamma  \frac {\left|V\right|}{M_p^2} \ .
  \label{buenaes}
 \eeq
 This constraint is represented in Fig. (\ref{barra2}).
 This gives a low-energy bound on the mass of the moduli at the minimum in terms of the value of the potential.
 It  is also somewhat analogous to the refined dS conjecture for $c'=\gamma$ and the recent TCC conjecture \cite{Bedroya:2019snp}, but it also applies for AdS vacua.
 
 In this case one can see that dS minima are forbidden, at least for potentials such that $V\rightarrow 0$ as $T$ goes to infinity. 
 The point is that,  in such a case, if the potential has a dS minimum, then it must necessarily have a dS local maximum. 
 If the potential is  non-singular, there should be a field path connecting the minimum with the maximum. 
 But, as the figure shows, one cannot continuously go from a minimum to a maximum without going through the
 forbidden region where gravity is too strong.
 
 It would be interesting to  test these minima conditions in the context of the class of Type IIA AdS 
 vacua in ref.\cite{DeWolfe:2005uu,Camara:2005dc,Aldazabal:2006up,Blumenhagen:2019vgj,Font:2019uva} .
 A different conjecture also involving also second derivatives of the potential was put forward in
 \cite{Andriot:2018mav}.

One should take these bounds on potentials with caution.
Here we are only considering  one  modulus with a single massive object verifying PPSWGC constraints,
and with a simple potential of the form $V\simeq m^{2\gamma}$. 
Still it shows how a possible connection between the dS  and scalar WGC conjectures could arise. 
It would be particularly  interesting to generalize, if possible,  these arguments to the
case of multiple fields.

 \section{Strong Scalar Weak Gravity Conjecture}
 
 The proposed  PPSWGC declares that WGC particles with mass $m$ must exist
 such that their production rate by massless moduli $\phi$  is constrained as in section 3.
  On the other hand, according to the idea that no interaction weaker than gravity exists, 
  one would also expect constraints among  interacting  scalars. In particular, moduli will in general
  acquire masses and interactions in the absence of SUSY, and their interactions would be constrained 
  if they must be weaker than gravity. In our previous paper, we named it the \emph{strong Scalar Weak Gravity Conjecture (sSWGC)}.
    Inspired by the PPSWGC and the scalar constraints  found in section 3 one may conjecture for the case of a single self-interacting 
 a sSWGC constraint of the form:
 \beq
\boxed{  \left| \xi(V''')^2 \ -\ (V'')(V'''')\right|\ \geq \ \frac {(V'')^2}{M_p^2}  }\ ,
\label{nuevastrong}
 \eeq
 where we have made the replacement $m^2\rightarrow V''$.
Here $\xi$ is some number of order one.
  In \cite{Gonzalo:2019gjp}  such a constraint was proposed with $\xi=2$, inspired by eq.(\ref{viejaseminal}). On the other hand 
  eq.(\ref{viejauno}) would suggest instead $\xi =1$.  These proposal was not based on a specific set of diagrams but are rather
  bold generalizations to the case of the self-interactions of a single real scalar field.
    
  One may try to  justify  a particular value of $\xi$  in terms of the scattering of a couple of real scalars.  In this case one would 
  consider the elastic scattering of two scalars  almost at rest compared to the threshold production of two scalars from graviton
  scattering. The diagrams for the scalar scattering are those as in Fig. (\ref{diagrama3}) and an additional $s$-channel one in which two scalars
  produce a virtual scalar which then decays into two scalars. Since the mediators are massive, the threshold conditions are $u=t=0$ and also $s=4m^2$. Direct application of the Feynman rules give:
  \beq
  N_{1}+N_{2}+N_{3}=-(\dot{m}^{2})^{2}\left\{ \frac{1}{t-m^{2}}+\frac{1}{u-m^{2}}+\frac{1}{s-m^{2}}\right\} -\ddot{m}^{2},
  \eeq
  so one obtains $\xi =5/3$.  
  Still, as we will see below, in order to see how masses and parameters 
  are constrained the particular value of $\xi$ is not very relevant, as long as it is of order one.

\begin{figure}[H]
	\centering{}
	\label{barra3}
	\includegraphics[scale=0.38]{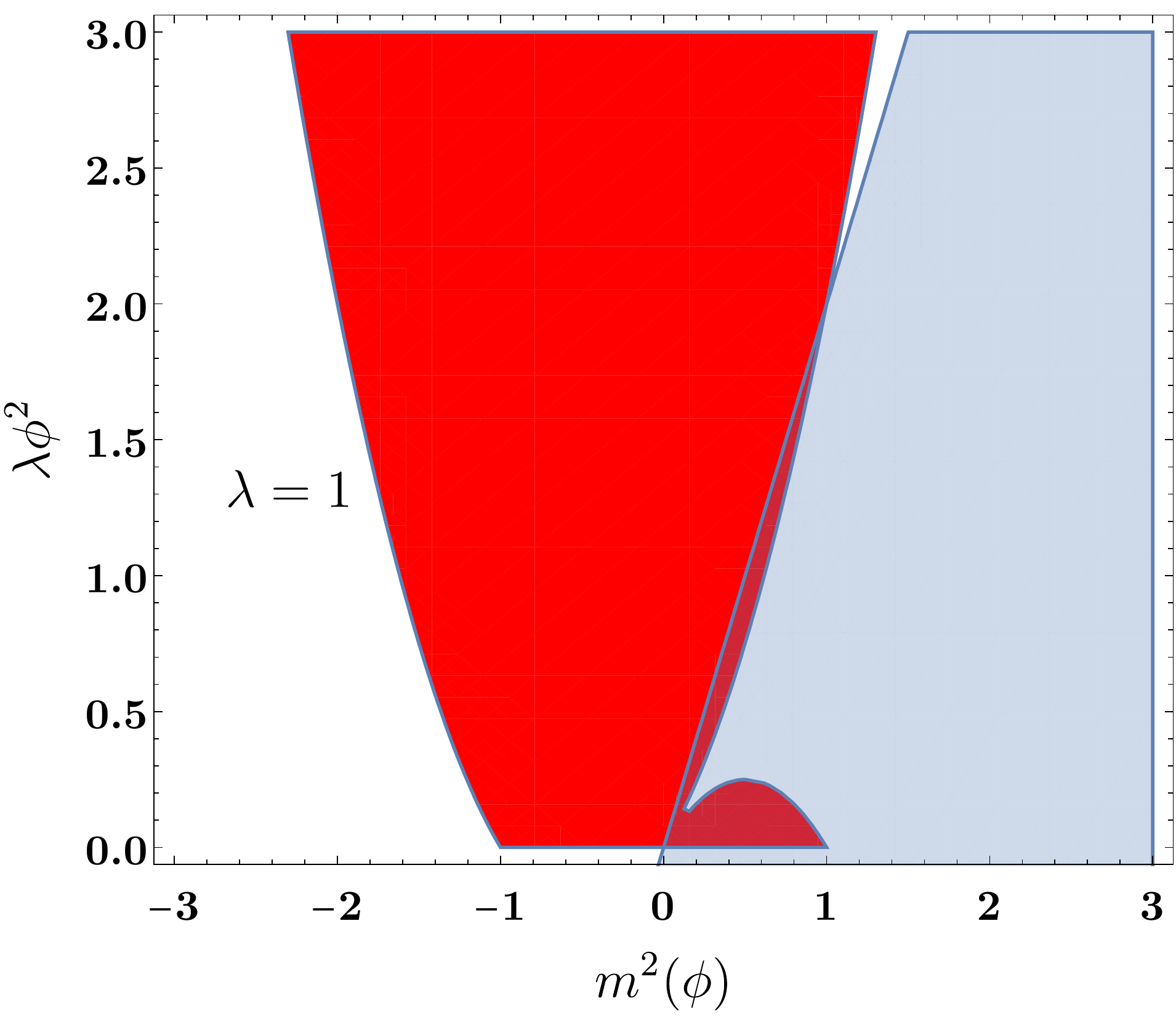}
	\includegraphics[scale=0.38]{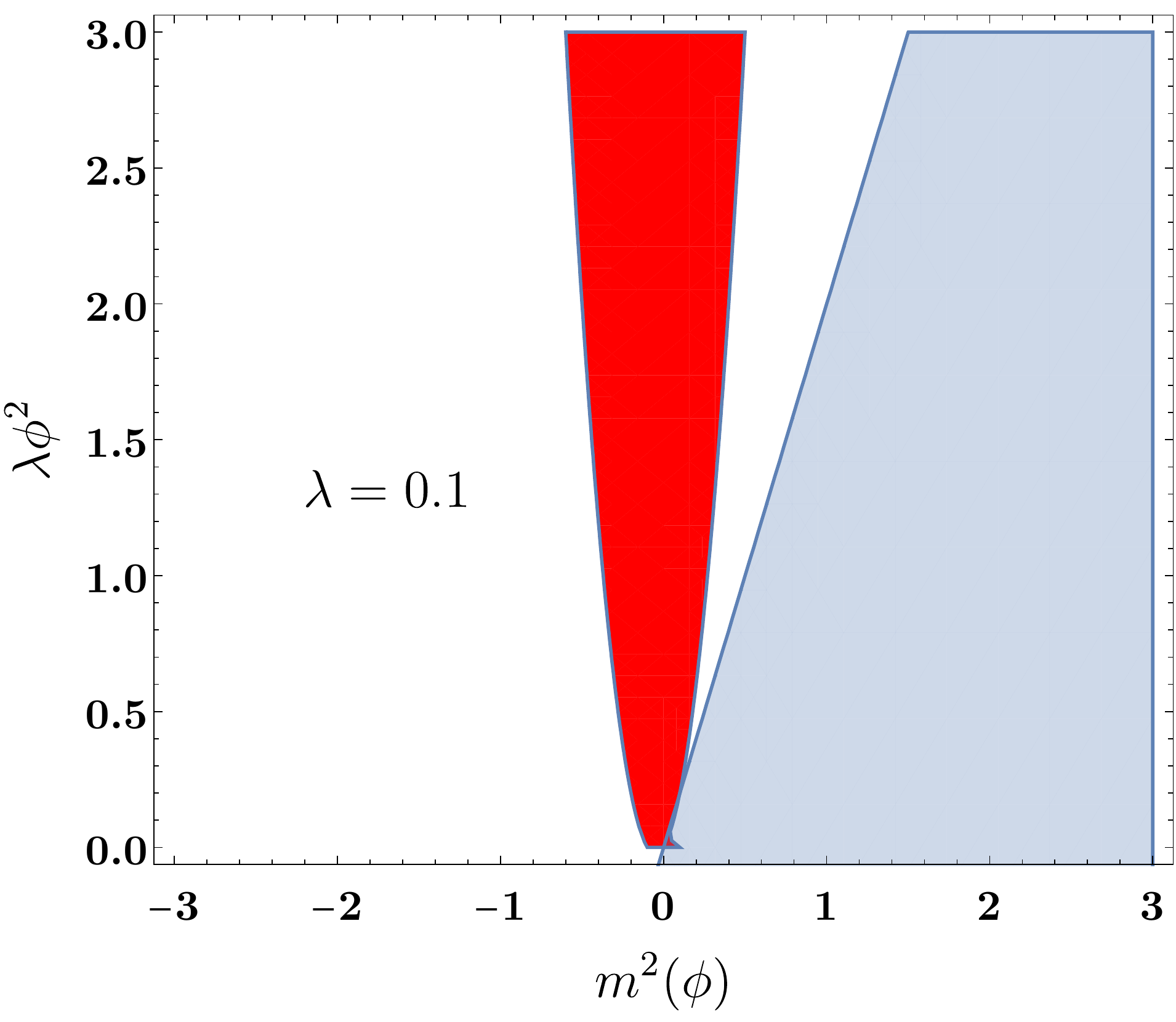}
\caption{\footnotesize Constraints  on the $m^2(\phi)-\lambda \phi^2$ plane for a quartic real scalar potential for fixed $\lambda=1$ and $0.1$ in 
$M_p/\xi^{1/2}$ units.
	The red region is allowed. The blue region marks when $m_0^2>0$.  As the field $\phi$ tends to zero, the field-dependent mass   $m^2(\phi)$ also decreases.
	The allowed region also gets narrower as $\lambda$ decreases.
	When gravity decouples the allowed red region covers all the upper half-plain. 
	}
\end{figure}

   One important property of the  condition  (\ref{nuevastrong})  is that it is a Swampland condition, in the sense that it disappears  when
 gravity decouples. This is unlike the condition without the absolute value put forward in \cite{Gonzalo:2019gjp}.
 The constraint in Eq. (\ref{nuevastrong}) passes some interesting tests. It is easy to check that an axion potential of the form 
 $V(\eta)=-M^4 \text{cos}(\eta/f)$ obeys the constraint as long as the decay constant
 obeys $f\leq M_p$, in agreement with axion WGC arguments \cite{Gonzalo:2019gjp}. If one considers a Higgs-like potential of the form
 $V=m_0^2\phi^2/2+\lambda_0 \phi^4/4!$ one gets from Eq. (\ref{nuevastrong}) a constraint
 %
 %
 \beq
 |\lambda | \left| m^2(\phi)\ -\ \lambda \phi^2 \right| \ \geq \  \xi \frac {m^4(\phi)}{M_p^2} \ ,
 \eeq
 where  $\lambda=\xi\lambda_0$ and $m^2=V''=m_0^2+\frac {1}{2}\lambda_0 \phi^2$ the field-dependent mass$^2$.
 Note that  at small $\phi$ the constraint is verified as long as $|\lambda_0|\geq (m_0/M_p)^2$, in agreement with  Weak Gravity intuitions. 
 The constraints on the plain $m^2(\phi)$-$\lambda \phi^2$ are shown in Fig. (\ref{barra3}) for  $\lambda=1, 0.1$, in units of $M_p/\xi^{1/2}$. The red area is 
 allowed by the condition. One notices that as the value of the field $\phi\rightarrow 0$ the value of the mass becomes smaller 
 and smaller. This pattern is sharper for smaller $\lambda$. This structure is  interesting, as it shows 
 an unexpected field-dependent upper bound on the masses of scalars as the field varies.  
  For values outside the red area the gravitational interaction dominates over the scalar interaction and hence that situation is forbidden.
 As gravity decouples, the red  area covers all the plain.

 The  sSWGC  without the absolute value, for the Higgs-like potential at $\phi=0$ would give  $\lambda \leq -\frac{m_0^2}{M_p^2}$. 
 Thus for a massive scalar only an unstable potential with $\lambda<0$ would be allowed.
  Based on this observation  (as applied to the case in \cite{Gonzalo:2019gjp})
 some  counter-examples were argued to exist in \cite{Freivogel:2019mtr}. In the new 
  results  in the present paper both signs of $\lambda$ are allowed.
 In this regard note that a constraint without the absolute value, would  also forbid field ranges with $|\phi|^2\leq 2m_0^2/\lambda$, even in the
 absence of gravity. Such a forbidden field range is no longer present in the new constraints Eq. (\ref{nuevastrong}). Note that somewhat analogous results, in particular, the presence of an absolute value and the coefficient $\frac{5}{3}$ were also obtained in \cite{Benakli:2020pkm}, although starting from different physical principles.
%

 
 As we said, as $\lambda\rightarrow 0$ the condition is violated.
  It was already pointed out in \cite{Gonzalo:2019gjp} that, since $\lambda$ in the SM vanish at a 
 high scale $\simeq 10^{10}-10^{13}$ GeV, new physics is predicted to appear at this scale, since the quartic interaction becomes weaker than gravity.
  An elegant solution to this problem is that SUSY 
 is recovered below that scale, getting a theory consistent with quantum gravity.
  Note that this behaviour  appears only in the presence of gravity and hence would be a Swampland constraint, not a field theory constraint.

 The possibility of a cancellation between trilinear and quartic contributions is more general than the above constraint, and 
 appear in other examples due to the structure of the amplitude in Eq.(\ref{scalarppwgc}). In fact this is the origin of the 
 forbidden bands in Figs. (\ref{barra1}),(\ref{barra2}).    This may lead to potential inconsistencies 
 with the scalar WGC at finite points in moduli space in some examples, indicating their inconsistency or incompleteness if
  gravity is present. In the $N=2$ SUSY examples shown above this does not happen for the BPS states, which lie at the boundary 
  of the forbidden regions.
  But  it may  happen e.g.  in non-SUSY examples for
  some field value.  Turning the argument around, the presence of these forbidden regions  in non-SUSY theories coupled to gravity 
  could  be an argument for the presence of SUSY at some scale in the low energy effective action.

It is important to remark that the sSWGC stands on a less firm ground than the general PPWGC or the SWGC discussed in sections 2 and 3.
In particular it is not obvious that the simple recipe 
 $m^2\rightarrow V''$ in the PPWGC constraints is sufficiently justified.
To our knowledge, there is however no counterexample to the sSWGC here considered with an absolute value included. It would be interesting to find further support for  generalized SWGC like this. If one takes a constraint like Eq.(\ref{nuevastrong}) to be valid for any single scalar potential, there are important phenomenological implications,  as already shown for the old version of the constraint in \cite{Gonzalo:2019gjp}.
 In addition, it would also be  interesting  to find a multi-field generalization of these constraints.

\section{Final comments and conclusions}

In this paper we have proposed pair production of massive particles at threshold as a means to 
compare the gravitational  to the gauge and scalar interactions. 
Equivalent results would be obtained from pair annihilation of massive particles into photons/gravitons/moduli.
Imposing that the production rates from 
gravitons is always smaller than that from gauge bosons or moduli gives rise to specific  WGC constraints.
In the case of $U(1)$ interactions this diagrammatic prescription reproduces the same results as obtained from
instability of extremal black-holes. On the other hand when applied to pair creation from  moduli,  a scalar 
WGC constraint depending on first and second derivatives of the mass appear. Intriguingly, imposing saturation of
the conditions one obtains simple differential equations. Some of the solutions match with known results 
in $N=2$ BPS examples and are consistent with the Swampland Distance Conjecture. Other solutions 
have more general asymptotic behaviour.

One interesting aspect of this approach  is that it derives the $U(1)^n$ WGC conjectures and 
a scalar WGC from the same general  principle of gravity being the weakest interaction. The form
of the scalar WGC depends on whether we are dealing with complex or real moduli and the
metric in moduli space. For the case of n complex moduli the constraint
 Eq. (\ref{varioscomplex}) is obtained.  One has to view our proposal as complementary to the 
constraints obtained from extremal black-hole instability and the Repulsive conjectures. We think our
proposal is particularly interesting in its application to obtain  constraints on scalar couplings.

  One point to note is that our condition is a quantum relativistic condition since it involves 
  particle production and interaction  rates rather than amplitudes. This is unlike the case of
  one photon/graviton exchange with particles at rest which give rise to the classical non-relativistic 
  Coulomb/Newton potentials. The presence of rates (absolute values of amplitudes) plays also an
  important role in the emergence  of duality symmetries among the states saturating the bounds.
It is  particularly remarkable how the existence  of momenta and winding (extended objects) emerges from
simple scattering amplitude considerations in the effective low-energy theory.

There are many aspects which deserve further study. One interesting  question is the applicability of
the constraints of scalar moduli in non-SUSY theories, in which a moduli space of massless moduli does
not in general exist. In this connection,  some of the  saturating solutions for the scalar WGC constraints
that we obtain 
may be interpreted as the bosonic subsector of BPS and special geometry conditions in $N=2$
supergravity theory. 
On the other hand we believe that the principle of gravity being the weakest force 
is independent of supersymmetry and one can expect that the constraints will  still apply at least in theories 
with  spontaneously broken SUSY.  

It would be interesting to test the condition Eq.(\ref{varioscomplex})   in specific string settings, like
the towers of BPS states getting massless at large moduli in Type IIA and Type IIB CY compactifications, as in
refs.\cite{Grimm:2018ohb,Corvilain:2018lgw,Font:2019cxq,timo,Gendler:2020dfp}. 
Another interesting direction for further research would include the extension to higher dimensions and
to non-Abelian gauge groups. It would also be important to obtain constraints on scalar  potentials of moduli and
other scalar fields along the lines of sections 4  (Eq. (\ref{buenaes})) and 5 in this paper and study its implications in cosmology and particle physics.

It is important to determine what is it exactly that goes wrong if the PPWGC condition is violated.  Pair production of charged particles is a characteristic of black-hole radiation and it would be important to elucidate the precise connection, if at all,  of the present ideas with black-hole physics. 
Our conditions also  imply that the annihilation rate of charged black-holes into photons must be larger than to gravitons.
Perhaps the PPWGC constraints appear as  additional  dynamical requirements.
It would be interesting to extend the results of this work by considering pair production of particles in backgrounds different from flat space-time, not only in the context of black-holes but also in $AdS$ and $dS$.

More generally, we would like to understand whether and why gravity should  be weaker than any other 
interaction, and the role of this condition in the general context of Quantum Gravity and String Theory.

\vspace{0.5cm}

\centerline{\bf \large Acknowledgments}

\bigskip

\noindent We thank G. Aldazabal, A. Font, A. Herr\'aez,  F. Marchesano, M. Montero,  A. Uranga, and I. Valenzuela for useful discussions. 
This workis  is supported  by  the  Spanish  Research  Agency  (Agencia  Espa\~nola  de  Investigaci\'on) through  the  grants  IFT  Centro  de  Excelencia  Severo  Ochoa  SEV-2016-0597, the grant GC2018-095976-B-C21from MCIU/AEI/FEDER, UE and the grant PA2016-78645-P.
   E.G. is supported by the Spanish FPU Grant No. FPU16/03985.

\end{document}